\documentclass{iopart}
\usepackage{euscript,iopams,amssymb,amsfonts,graphicx,bm}
\usepackage{pgfplots,setstack}

\usepackage{float}
\usepackage{epsfig}
\usepackage{esint}
 \usepackage{tikz-cd}

\usepackage{bm,braket}

\eqnobysec

\newcommand{\wrho}{\widetilde{\rho}}

\newcommand{\J}{\widetilde{J}}

\newcommand{\G}{\widetilde{G}}
\newcommand{\K}{\widetilde{K}}
\newcommand{\calU}{{\mathcal U}}

\newcommand{\calM}{{\mathcal M}}

\newcommand{\R}{{\mathbb R}}
\renewcommand{\L}{{\mathbb L}}
\newcommand{\X}{\mathbf{X}}

\renewcommand{\P}{\mathbb{P}}
\newcommand{\PP}{\widetilde{P}}
\newcommand{\p}{\widetilde{p}}
\newcommand{\f}{\widetilde{f}}

\newcommand{\q}{\widetilde{q}}
\newcommand{\Q}{\widetilde{Q}}
\newcommand{\x}{\mathbf{x}}
\newcommand{\y}{\mathbf{y}}
\newcommand{\z}{\mathbf{z}}
\renewcommand{\e}{{\mathrm e}}
\newcommand{\E}{{\mathbb E}}

\newcommand{\n}{\mathbf n}
\newcommand{\calT}{{\mathcal T}}
\renewcommand{\P}{\mathbb P}

\renewcommand{\S}{\widetilde{S}}

\begin{document}

 \title[Renewal theory for Brownian motion across a stochastically gated interface]{Renewal theory for Brownian motion across a stochastically gated interface}

\author{Paul C. Bressloff}
\address{Department of Mathematics, Imperial College London, 
London SW7 2AZ, UK}
\ead{p.bressloff@imperial.ac.uk}

\begin{abstract} Stochastically gated interfaces play an important role in a variety of cellular diffusion processes. Examples include intracellular transport via stochastically gated ion channels and pores in the plasma membrane of a cell, intercellular transport between cells coupled by stochastically gated gap junctions, and oxygen transport in insect respiration. Most studies of stochastically-gated interfaces are based on macroscopic models that track the particle concentration averaged with respect to different realisations of the gate dynamics. In this paper we use renewal theory to develop a probabilistic model of single-particle Brownian motion (BM) through a stochastically gated interface. (An analogous theory for semipermeable interfaces was previously developed using so-called snapping out BM.) We proceed by constructing a renewal equation for 1D BM with an interface at the origin, which effectively sews together a sequence of BMs on the half-line with a totally absorbing boundary at $x=0$. Each time the particle is absorbed, the stochastic process is immediately restarted according to the following rule: if the gate is closed then BM restarts on the same side of the interface, whereas if the gate is open then BM restarts on either side of the interface with equal probability. In order to ensure that diffusion restarts in a state that avoids immediate re-absorption. we assume that whenever the particle reaches the interface it is instantaneously shifted a distance $\epsilon$ from the origin. We explicitly solve the  renewal equation for $\epsilon>0$ and show how the solution of a corresponding forward Kolmogorov equation is recovered in the limit $\epsilon\rightarrow 0$. However, the renewal equation provides a more general mathematical framework for modelling a stochastically gated interface by explicitly separating the first passage time problem of detecting the gated interface (absorption) and the subsequent rule for restarting BM. We conclude by discussing some of the mathematical challenges in extending the theory to higher-dimensional interfaces.

\end{abstract}

\maketitle


\section{Introduction}

Diffusion processes in heterogeneous media partitioned by semipermeable membranes or interfaces have extensive applications in both natural and artificial systems.  Examples include multilayer electrodes and semi-conductors \cite{Freger05,Gurevich09}, thermal conduction in composite media \cite{Barbaro88,Grossel98,deMonte00,Lu05}, diffusion magnetic resonance imaging (dMRI) \cite{Tanner78,Callaghan92,Coy94,Grebenkov10,Moutal19} and drug delivery \cite{Pontrelli07,Todo13,Farago18}. A classical example of a semi-permeable membrane in cell biology is a lipid bilayer that regulates the flow of macromolecules and ions between the intracellular and extracellular environments  or between subcellular compartments \cite{Alberts15,Bressloff22B,Nik21}. Analogous processes occur at the multicellular level, where small nonselective channels known as gap junctions allow the diffusive transport of molecules between neighbouring cells \cite{Brink85,Ramanan90,Good09}.
At the macroscopic level, multi-particle diffusion across a semi-permeable membrane is modelled by taking the Fickian flux across the membrane to be continuous and to be proportional to the difference in concentrations on either side of the barrier; the constant of proportionality is identified as the permeability. For example, consider one-dimensional diffusion with a semi-permeable barrier at $x=0$.  Let $u(x,t)$ be the concentration of particles at $x$ at time $t$. Then $u(x,t)$ is the solution of the diffusion equation with semipermeable or leather boundary conditions at $x=0$:
\numparts
\begin{eqnarray}
\label{classa}
\frac{\partial u(x,t)}{\partial t}&=D\frac{\partial^2 u(x,t)}{\partial x^2}, \quad x\neq 0,\\
J(0^{\pm},t)&=\kappa_0[u(0^-,t)-u(0^+,t)],
\label{classb}
\end{eqnarray}
\endnumparts
where $J(x,t)=-D\partial_xu(x,t)$ is the Fickian flux, $D$ is the diffusivity, and $\kappa_0$ is the (constant) permeability. Moreover, we define $f(0^{\pm},t)=\lim_{a \rightarrow 0}f(\pm a,t)$ with $a >0$ for any piecewise continuous function $f(x,t)$. 

A second type of barrier to diffusion prevalent in cellular transport processes is a stochastically gated interface. Examples include intracellular transport via stochastically gated ion channels and pores in the plasma membrane of a cell \cite{Reingruber10,Bressloff15a,Bressloff15b,Godec17}, intercellular transport between cells coupled by stochastically gated gap junctions  \cite{Bressloff16a,Bressloff16b}, and oxygen transport in insect respiration \cite{Lawley15b,Berez16}. In the case of 1D diffusion, suppose that the interface at $x=0$ stochastically switches between a closed (impermeable) state $\sigma(t)=-1$ and an open (permeable) state $\sigma(t)=1$ according to a two-state Markov chain with transitions
\[-1\overset{\beta}\longrightarrow 1,\quad 1\overset{\alpha}\longrightarrow  -1. \]
Let $U(x,t)$ denote the concentration of particles at time $t$, which evolves according to the stochastic hybrid diffusion equation
\numparts
\begin{eqnarray}
\label{sclassa}
&\frac{\partial U(x,t)}{\partial t}=D\frac{\partial^2 U(x,t)}{\partial x^2}, \quad x\neq 0,\\
&(1-\sigma(t))\partial_x U(0,t)=0.
\label{sclassb}
\end{eqnarray}
\endnumparts
That is, there is a no-flux condition at $x=0$ when the gate is closed. Introducing the averaged concentrations
\begin{equation}
u_k(x,t)=\E[U(x,t)\delta_{\sigma(t),k}],\quad k=\pm 1,
\end{equation}
where expectations are taken with respect to realisations of the switching process, it can be shown that $u_k$ satisfies the deterministic forward Kolmogorov equation \cite{Lawley15b,Bressloff15a,Lawley15a}
\numparts
\begin{eqnarray}
\label{1DUa}
 &\frac{\partial u_1(x,t)}{\partial t}=D\frac{\partial^2u_1(x,t)}{\partial x^2}-\alpha  u_1(x,t)+\beta u_{-1}(x,t),\ x\neq 0,\\
 \label{1DUb}
 & \frac{\partial u_{-1}(x,t)}{\partial t}=D\frac{\partial^2u_{-1}(x,t)}{\partial x^2}+\alpha u_1(x,t)-\beta u_{-1}(x,t),\ x\neq 0,\\
\label{1DUc}
&\frac{\partial u_{-1}(0^+,t)}{\partial x}=0,\quad \frac{\partial u_{-1}(0^-,t)}{\partial x}=0,\\ 
\label{1DUd}
&u_1(0^+,t)=u_1(0^-,t),\quad \frac{\partial u_1(0^+,t)}{\partial x}=\frac{\partial u_1(0^-,t)}{\partial x}.
\end{eqnarray}
\endnumparts

Note that there is a link between semi-permeable and stochastically gated interfaces as shown in Ref. \cite{Lawley15a}. These authors prove analytically that equations (\ref{classa}) and (\ref{classb}) can be recovered by setting $u(x,t)=u_1(x,t)+u_{-1}(x,t)$ and taking the double limit $\alpha\rightarrow \infty$ and $\beta \rightarrow \infty$ under the additional constraint that the proportion of time that the gate spends in the closed state approaches unity.
More precisely, introduce a fixed fundamental rate $\gamma_0$ and scale the transition rates $\alpha,\beta$ as
\begin{equation}
\label{scale}
\alpha = \frac{ (1-\eta)\gamma_0 }{\delta},\quad \beta =\frac{\eta \gamma_0 }{\delta},
\end{equation}
with $0<\eta <1$ and $0<\delta \ll 1$. The boundary condition (\ref{classb}) with permeability $\kappa_0$ is then obtained in the limit $\delta \rightarrow 0$ if \cite{Lawley15a}
\begin{equation}
\label{eta}
\eta\sim  2\kappa_0\sqrt{\frac{\delta}{\gamma_0 D}}.
\end{equation}
In other words, $\alpha = O(1/\delta)$ and $\beta=O(1/\sqrt{\delta})$ as $\delta \rightarrow 0$.

Recently, a probabilistic model of single-particle diffusion across a semi-permeable membrane has been developed based on so-called snapping out Brownian motion (BM) \cite{Lejay16,Aho16,Lejay18,Farago20,Brobowski21,Bressloff22,Bressloff23a,Bressloff23b}. The latter effectively sews together partially reflected BMs on either side of an impermeable interface to obtain the corresponding stochastic dynamics across a semi-permeable interface. Each round of partially reflected BM in $[0,\infty)$ or $(-\infty,0]$ is terminated when the local time at $x=0^+$ or $x=0^-$ exceeds an exponential random variable parameterized by $\kappa_0$\footnote{For reflected Brownian motion in $[0,\infty)$, the local time is defined according to $L_+(t)=D\lim_{a \rightarrow 0}a^{-1} \Theta(a -X(t))$, where $X(t)$ is the position of the particle at time $t$ and $\Theta(x)$ is a Heaviside function. A sample path is terminated at the stopping time $\calT=\inf\{t>0,\ L_+(t)>h\}$ where $h$ is independently identically distributed according to the exponential distribution $(\kappa_0/D)\e^{-\kappa_0/D}$. Similarly, the local time for reflrcted BM in $(-\infty,0]$ is $L_-(t)=D\lim_{a \rightarrow 0}a^{-1} \Theta(a +X(t))$.}. Following each termination event, the particle immediately resumes either negatively or positively reflected BM with equal probability. Moreover, one can construct a renewal equation that relates the probability density $\rho(x,t)$ for samples paths of snapping out BM to the corresponding probability density $p(x,t)$ for Brownian motion on the half-line with a Robin boundary condition at $x=0$ \cite{Bressloff22,Bressloff23a}. This then establishes that $\rho(x,t)$ satisfies a semi-permeable boundary condition of the form (\ref{classb}).

In this paper, we develop an analogous renewal formulation of Brownian motion across a stochastically gated interface by adapting recent work on first passage time (FPT) problems for stochastically gated targets \cite{Bressloff26}.
We begin in section 2 by considering BM in $\R$ with a stochastically gated interface at $x=0$, and analyse the forward Kolmogorov equation for the probability densities $\rho_k(x,t)=\E[\rho(x,t)\delta_{k,\sigma(t)}]$, where $\rho(x,t)$ is the marginal probability density for particle position. In section 3, we construct a first renewal equation that relates $\rho_k(x,t)$ to the corresponding probability density $p(x,t)$ for BM on the half-line with a totally absorbing boundary condition at $x=0$. The renewal equation on $\R$ effectively sews together a sequence of BMs on the half-line in an analogous fashion to snapping out BM \cite{Bressloff22,Bressloff23a}. Each time the particle is absorbed, the stochastic process is immediately restarted according to the following rule: if the gate is closed then BM restarts on the same side of the interface, whereas if the gate is open then BM continues freely on either side of the interface. However, as previously found for stochastically gated targets \cite{Bressloff26}, a crucial step in sewing together successive rounds of BM on the half-line is ensuring that diffusion restarts in a state that avoids immediate re-absorption. (This is straightforward in the case of snapping out BM because each side of the interface is only partially absorbing.) Therefore, following along analogous lines to Ref. \cite{Bressloff26}, we assume that whenever the particle reaches the interface, it is instantaneously shifted a distance $\epsilon$ from the origin. The latter can be interpreted as a form of so-called boundary-induced resetting \cite{Bressloff25b,Bressloff26}. We then obtain an explicit solution of the  renewal equation for $\epsilon>0$ using Laplace transforms and the convolution theorem. A non-trivial step in the calculation is deriving self-consistency conditions for the solution under the initial conditions $x=\pm \epsilon$. We then show how the resulting solution in the limit $\epsilon\rightarrow 0$ is equivalent to the solution of the Kolmogorov equation obtained in section 2. However, the renewal equation provides a more general mathematical framework for modelling a stochastically gated interface by explicitly separating the FPT problem for detecting the gated interface (absorption) and the subsequent rule for restarting BM.

In section 4 we extend the renewal equation to include the effects of stochastic resetting in the bulk domain. That is, during each round of BM on the positive (negative) half-line prior to absorption at the interface, the particle can reset to a fixed location $x=\xi$ ($x=-\xi$) at a constant Poisson rate $r$. One of the characteristic features of non-absorbing diffusion processes with stochastic resetting is that there exists a nonequilibrium stationary state (NESS), which is maintained by non-zero probability fluxes \cite{Evans11a,Evans11b,Evans20}. Although each round of BM on the half-line is terminated by absorption at the interface, the stochastic process is immediately restarted so that BM across the stochastically gated interface is not terminated and supports an NESS. We use the renewal equation to calculate the NESS in the case of both spontaneous and boundary-induced resetting, and determine its dependence on the distance $\epsilon$. Finally, in section 5, we indicate some of the mathematical challenges in extending the renewal approach to higher-dimensional interfaces.

\section{Stochastically gated interface in $\R$}

\begin{figure}[t!]
  \centering
  \includegraphics[width=10cm]{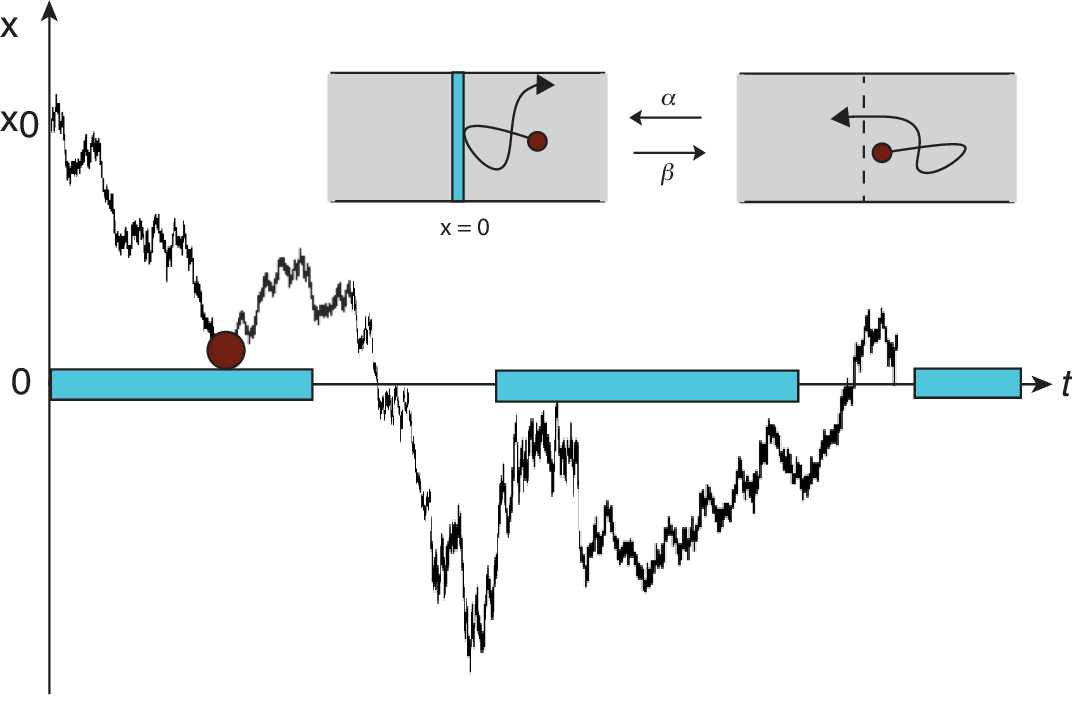}
  \caption{Brownian particle in $\R $ with a stochastically gated interface at $x=0$. The gate is open (permeable) if $\sigma(t)=1$ and closed (impermeable) if`$\sigma(t)=-1$. Transitions $\sigma(t)\rightarrow -\sigma(t)$ occurs according to a two-state Markov chain with transition rates $\alpha,\beta$.}
  \label{fig1}
\end{figure}

  Consider a Brownian particle diffusing in $\R$ with a stochastically gated barrier or interface at $x=0$, see Fig. \ref{fig1}. Let $X(t)\in \R$ denote the position of the particle at time $t$ and $\sigma(t)\in \{1,-1\}$ the corresponding state of the gate. The gate is open (permeable) if $\sigma(t)=1$ and closed (impermeable) otherwise. We assume that the discrete process $\sigma(t) $ evolves according to a two-state Markov chain with transition matrix
  \begin{equation}
{\bf K}=\left (\begin{array}{cc} 0& \beta \\ \alpha & 0 \end{array} \right ).
\label{axe2}
\end{equation}
 If $\Pi_{kk_0}(t)=\P[\sigma(t)=k|\sigma(0)=k_0]$, then the master equation for $\sigma(t)$ takes the form
\begin{equation}
\frac{d\Pi_{k k_0}}{dt}=\sum_{m=\pm 1}\bigg [K_{km}-\delta_{k,m}\sum_{l=\pm 1} K_{lk} \bigg ]\Pi_{mk_0},\quad k=\pm 1, \end{equation}
with $\Pi_{-1k_0}(t)+\Pi_{1k_0}(t)=1$. It follows that
\begin{equation}
\label{Pi}
\Pi_{kk_0}(t)=\delta_{k,k_0}\e^{-t/\tau_c}+\sigma_k (1-\e^{-t/\tau_c}),\quad \tau_c=\frac{1}{\alpha+\beta},
\end{equation}
with $\sigma_1=\beta/(\alpha+\beta) $ and $\sigma_{-1}=\alpha/(\alpha+\beta) $.
Here $\tau_c$ is the relaxation time such that $\Pi_{kk_0}(t)\rightarrow \sigma_k$ in the limit $t\rightarrow \infty$. 
Let $\rho_{k|k_0}(x,t|x_0)$ denote the full propagator for the pair $(X(t),\sigma(t))$ with
\[\fl \rho_{k|k_0}(x,t|x_0)dx=\P[x<X(t)<x+dx,\sigma(t)=k|X(0)=x_0,\sigma(0)=k_0]\]
and the initial condition $\rho_{k|k_0}(x,0|x_0)=\delta(x-x_0)\delta_{k.k_0}$. For concreteness, we will assume that $x_0>0$.
Let $\Omega=(-\infty,0^-]\cup [0^+,\infty)$. The propagator evolves according to the forward Kolmogorov equation
\numparts
\begin{eqnarray}
\label{1Da}
\fl &\frac{\partial \rho_{1|k_0}(x,t|x_0)}{\partial t}=D\frac{\partial^2\rho_{1|k_0}(x,t|x_0)}{\partial x^2}-\alpha  \rho_{1|k_0}(x,t|x_0)+\beta \rho_{-1|k_0}(x,t|x_0),\\
 \label{1Db} 
 \fl & \frac{\partial \rho_{-1|k_0}(x,t|x_0)}{\partial t}=D\frac{\partial^2\rho_{-1|k_0}(x,t|x_0)}{\partial x^2}+\alpha  \rho_{1|k_0}(x,t|x_0)-\beta \rho_{-1|k_0}(x,t|x_0),
\end{eqnarray}
 for all $x\in \Omega$,
\begin{eqnarray}
\label{1Dc}
\fl & \frac{\partial \rho_{-1|k_0}(0^+,t|x_0)}{\partial x}=0,\quad \frac{\partial \rho_{-1|k_0}(0^-,t|x_0)}{\partial x}=0,\\ 
\label{1Dd}
\fl &\rho_{1|k_0}(0^+,t|x_0)=\rho_{1|k_0}(0^-,t|x_0),\quad \frac{\partial \rho_{1|k_0}(0^+,t|x_0)}{\partial x}=\frac{\partial \rho_{1|k_0}(0^-,t|x_0)}{\partial x}.
\end{eqnarray}
\endnumparts
The interfacial boundary conditions at $x=0$ can be interpreted as follows. If the gate is closed then the particle is reflected by the barrier irrespective of whether it approaches from the right-hand side ($x=0^+$) or the left-hand side $(x=0^-)$. On the other hand, if the gate is open then the particle freely crosses from one side of the barrier to the other, so that the density and fluxes are continuous across $x=0$.

 \subsection{Solution in Laplace space}
Laplace transforming equations (\ref{1Da})--(\ref{1Dd}) gives
\numparts
\begin{eqnarray}
\label{1DLTa}
 \fl & D\frac{\partial^2\wrho_{1|k_0}(x,s|x_0)}{\partial x^2} -(\alpha+s)\wrho_{1|k_0}(x,s|x_0)+\beta  \wrho_{-1|k_0}(x,s|x_0)= -\delta(x-x_0)\delta_{1,k_0},\\
\fl   &D\frac{\partial^2\wrho_{-1|k_0}(x,s|x_0)}{\partial x^2} -(s+\beta)\wrho_{1|k_0}(x,s|x_0)+\alpha  \wrho_{1|k_0}(x,s|x_0)= -\delta(x-x_0)\delta_{-1,k_0},\nonumber \\
\fl 
 \label{1DLTb}
  \end{eqnarray}
  for $x\in \Omega$ and the interfacial conditions
  \begin{eqnarray}
  \label{1DLTc}
 \fl &\frac{\partial \wrho_{-1|k_0}(0^+,s|x_0)}{\partial x}=0,\quad \frac{\partial \wrho_{-1|k_0}(0^-,s|x_0)}{\partial x}=0\\
\fl & \wrho_{1|k_0}(0^+,s|x_0)=\wrho_{1|k_0}(0^-,s|x_0),\quad \frac{\partial \wrho_{1|k_0}(0^+,s|x_0)}{\partial x}=\frac{\partial \wrho_{1|k_0}(0^-,s|x_0)}{\partial x}.
\label{1DLTd}
\end{eqnarray}
\endnumparts
Adding equations (\ref{1DLTa}) and (\ref{1DLTa}), and setting $\rho(x,s)=\wrho_{1|k_0}(x,s|x_0)+\wrho_{-1|k_0}(x,s|x_0)$, we have  
\numparts\begin{eqnarray}
 & D\frac{\partial^2\rho(x,s)}{\partial x^2} -s\rho(x,s)= -\delta(x-x_0),\quad x>0,\\
 & D\frac{\partial^2\rho(x,s)}{\partial x^2} -s\rho(x,s)= 0, \quad x<0.
 \end{eqnarray}
 \endnumparts
Take the boundary condition for $\rho$ to be $ \rho(0^{\pm},s)=A_{\pm}(s)$ with $A_{\pm}(s)$ to be determined self-consistently. It follows that
\begin{eqnarray}
\label{q}
\rho(x,s)=\left \{ \begin{array}{cc}\G(x,s|x_0)+ \e^{-\sqrt{s/D}x}   A_+(s),\ &  x>0\\  \e^{\sqrt{s/D}x}   A_-(s),& x<0\end{array},\right .
\end{eqnarray}
where
\begin{equation}
\label{DG}
\G(x,s|x_0)=\frac{1}{2\sqrt{sD}}\bigg (\e^{-\sqrt{s/D}|x-x_0|}-\e^{-\sqrt{s/D}(x+x_0)}\bigg )
 \end{equation}
 is the Dirichlet Green's function of the modified Helmholtz equation on the half-line.  
 Given the solution $\rho(x,s)$ we set
 \begin{eqnarray}
 \fl \wrho_{1|k_0}(x,s|x_0)&=\sigma_{1}\rho(x,s)+j_{k_0}(x,s),\quad 
 \wrho_{-1|k_0}(x,s|x_0)=\sigma_{-1}\rho(x,s)-j_{k_0}(x,s),
 \end{eqnarray}
 with $j_{k_0}(0^{\pm},s)=B_{\pm}(s)$ and $B_{\pm}(s)$ to be determined self-consistently.
 Substituting into equation (\ref{1DLTa}) implies that
\numparts
 \begin{eqnarray}
 \fl & D\frac{\partial^2j_{k_0}(x,s)}{\partial x^2} -(\alpha+\beta +s)j_{k_0}(x,s)=-\delta(x-x_0)\bigg [\delta_{1,k_0}-\sigma_{1}\bigg ],\quad x>0,\\
 \fl & D\frac{\partial^2j_{k_0}(x,s)}{\partial x^2} -(\alpha+\beta +s)j_{k_0}(x,s)=0,\quad x<0.
 \end{eqnarray}
 \endnumparts
 The solution is of the form
 \begin{eqnarray}
 \label{j}
\fl j_{k_0}(x,s)=\left \{ \begin{array}{cc} k_0\sigma_{-k_0}\G(x,s+\alpha+\beta|x_0)+ \e^{-\sqrt{(s+\alpha+\beta)/D}x} B_+(s), & x>0,\\
 \e^{\sqrt{(s+\alpha+\beta)/D}x} B_-(s), & x<0 \end{array}.\right .
\end{eqnarray}

We have four unknown coefficients $A_{\pm}(s),B_{\pm}(s)$ that also depend on $(x_0,k_0)$ and four linearly independent boundary conditions given by (\ref{1DLTc}) and (\ref{1DLTd}). Expressing the latter in terms of $q$ and $j_{k_0}$, we have
\numparts
 \begin{eqnarray}
\fl  & \frac{\partial \rho(x,s)}{\partial x}|_{x=0^+} = \frac{\partial \rho(x,s)}{\partial x}|_{x=0^-},\quad  \frac{\partial j_{k_0}(x,s)}{\partial x}|_{x=0^+} = \frac{\partial j_{k_0}(x,s)}{\partial x}|_{x=0^-},\\
\fl  &\sigma_{-1}\frac{\partial \rho(x,s)}{\partial x}|_{x=0^-}=\frac{\partial j_{k_0}(x,s)}{\partial x}|_{x=0^-},\\
\fl &  \sigma_{1}\rho(0^+,s)+j_{k_0}(0^+,s)= \sigma_{  1}\rho(0^-,s)+j_{k_0}(0^-,s).
 \end{eqnarray}
 \endnumparts
 Combing with the general solutions (\ref{q}) and (\ref{j}) gives the four relations
  \numparts
  \begin{eqnarray}
  \label{conda}
 &h(x_0,s)=A_+(s)+A_-(s),\\
 \label{condb}
&k_0\sigma_{-k_0}h(x_0,s+\alpha+\beta)=B_+(s)+B_-(s),\\
  \label{condc}
  &\sigma_{-1}A_-(s)=\sqrt{\frac{s+\alpha+\beta}{s}}B_-(s),\\
\label{condd}
& \sigma_{1}A_+(s)+B_+(s)=\sigma_{1}A_-(s)+B_-(s).
  \end{eqnarray}
  \endnumparts
  We have set
  \begin{equation}
  h(x_0,s)= \frac{\e^{-\sqrt{s/D}x_0}}{\sqrt{sD}}.
  \end{equation}
  After some algebra, we obtain the following expressions for the coeeficients:
  \numparts
  \begin{eqnarray}
   \label{coefa}
\fl & A_+(s)=h(x_0,s)-A_-(s),\quad B_+(s)=k_0\sigma_{-k_0}h(x_0,s+\alpha+\beta)-B_-(s),\\
 \label{coefb}
\fl &  A_-(s)=\frac{1}{2}\frac{\sigma_1h(x_0,s)+k_0\sigma_{-k_0}h(x_0,s+\alpha+\beta)}{\sigma_1+\sqrt{s/(s+\alpha+\beta)}\sigma_{-1}},\\
 \label{coefc}
  \fl & 
   B_-(s)=\frac{\sigma_{-1}}{2}\frac{\sigma_1h(x_0,s)+k_0\sigma_{-k_0}h(x_0,s+\alpha+\beta)}{\sqrt{(s+\alpha+\beta)/s}\sigma_1+\sigma_{-1}}.
  \end{eqnarray}
  \endnumparts
 In summary, given these coefficients, the solution of equations (\ref{1DLTa})-- (\ref{1DLTd}) is
\numparts
\begin{eqnarray}
\label{sola}
\fl &\wrho_{k|k_0}(x,s|x_0) =\sigma_{k}\bigg [\G(x,s|x_0)+\e^{-\sqrt{s/D}x}A_+(s)\bigg ]\\
\fl& \hspace{2cm}   +k\bigg [k_0\sigma_{-k_0}\G(x,s+\alpha+\beta|x_0)+\e^{-\sqrt{[s+\alpha+\beta]/D}x}B_+(s)\bigg ],\quad x>0,\nonumber  \\
\fl  &\wrho_{k|k_0}(x,s|x_0) =\sigma_{k}\e^{\sqrt{s/D}x}A_-(s)+k\e^{\sqrt{[s+\alpha+\beta]/D}x}B_-(s),\quad x<0.
\label{solb}
\end{eqnarray}
\endnumparts

\subsection{Permeability in the fast switching limit}
The solution given by equations (\ref{sola}) and (\ref{solb}) will be used to validate the renewal approach in section 3. Here we show how the single-particle version of the semipermeable boundary condition (\ref{classb}) can be derived in the fast switching limit. We proceed using a similar scaling scheme to Ref. \cite{Lawley15a}.
First note that
\begin{equation}
\J(s)=-D \frac{\partial \rho(x,s)}{\partial x}|_{x=0^{\pm}} = -\sqrt{sD}A_-(s)
\end{equation}
and
\begin{equation}
\rho(0^-,s)-\rho(0^+,s)=A_-(s)-A_+(s).
\end{equation}
Introduce the scalings (\ref{scale}) with $0<\eta <1$ and $\gamma_0$ a baseline transition rate. Ignoring short-time transients, we can take the Laplace variable $s$ to be bounded from above and $\delta$ to be sufficiently small so that $\alpha+\beta \gg s$. 
Equation (\ref{condb}) implies that $B_+(s)\sim  -B_-(s)$ in the limit $\delta \rightarrow 0$. It then follows from equations (\ref{condc}) and (\ref{condd}) that
\begin{equation}
A_-(s)-A_+(s)\sim -\frac{2B_-(s)}{\sigma_1}\sim -\frac{2\sqrt{[\alpha+\beta]/D}}{\beta } \bigg [\sqrt{sD}A_-(s)\bigg ]
\end{equation}
in the limit $\delta \rightarrow 0$.
Substituting for $\alpha$ and $\beta$ using equation (\ref{scale}) gives
\begin{equation}
\rho(0^-,s)-\rho(0^+,s) \sim \frac{2}{\eta}\frac{\sqrt{\delta}}{\sqrt{ \gamma_0D}}    \J(s).
\end{equation}
Hence, we can take the limit $\delta \rightarrow 0$ to obtain a finite permeability $\kappa_0$ provided that $\eta$ satisfies equation (\ref{eta}).

\section{Renewal equation} 

 \begin{figure}[b!]
  \centering
  \includegraphics[width=10cm]{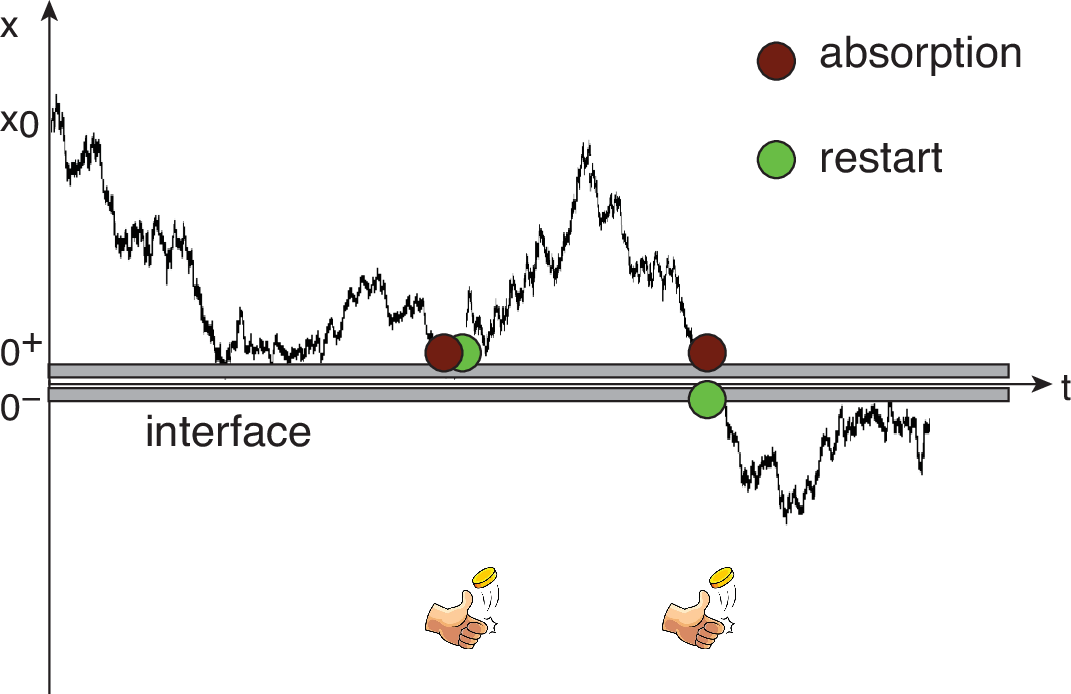}
  \caption{Snapping out BM in $\R $ with a semipermeable interface at $x=0$. Each side of the barrier is treated as a partially reflecting boundary. Schematic illustration of a sample trajectory starting at $x_0>0$. Each time the particle is absorbed at $x=0^{\pm}$ an unbiased coin is tossed to determine which side of the barrier BM restarts.}
  \label{fig2}
\end{figure}

\begin{figure}[t!]
  \centering
  \includegraphics[width=10cm]{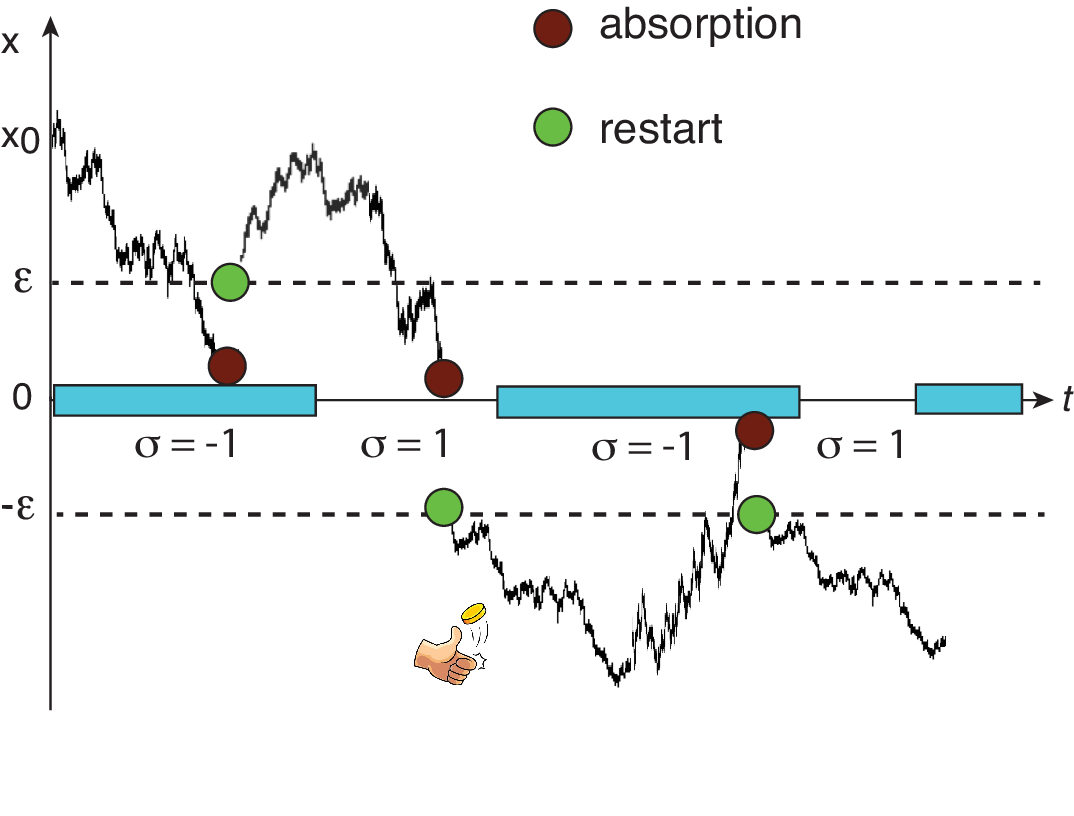}
  \caption{Brownian particle in $\R $ with a stochastically gated interface at $x=0$. Each side of the barrier is treated as a totally absorbing boundary. Schematic illustration of a sample trajectory starting at $x_0>0$. Each time the particle reaches $x=0$ and the gate is closed, BM restarts on the same side of the barrier and the restart position is taken to be a distance $\epsilon$ from the interface. On the other hand, if the gate is open then BM restarts on either side of the interface with equal probability.}
  \label{fig3}
\end{figure}

In this section we reformulate the stochastic process analysed in section 2 in terms of a renewal equation by extending our recent work on stochastically gated targets \cite{Bressloff26}. We begin by briefly recalling our previous renewal formulation of snapping out BM \cite{Bressloff22,Bressloff23a}. In the latter case, each side of the interface is treated as a partially reflecting boundary for Brownian motion,  which is implemented by imposing a Robin boundary condition for the corresponding forward Kolmogorov equation on the half line. Each time the particle is absorbed, the stochastic processes is immediately restarted with equal probability from either side of the interface, see Fig. \ref{fig2}. The corresponding renewal equation then sews together the resulting sequence of partially reflected BMs. There are two major differences in the renewal formulation of a stochastically gated interface. First, each side of the interface is taken to be totally absorbing so that as soon as the Brownian particle reaches $x=0$ the current round of diffusion is halted. Hence, the corresponding forward Kolmogorov equation on the half line has a Dirichlet boundary condition at $x=0$. Second, deciding which side of the barrier BM recommences now depends on the current state of the gate. That is, if the gate is closed then BM restarts on the same side of the interface, whereas if the gate is open then BM restarts on either side of the interface with equal probability, see Fig. \ref{fig3}. In order to handle the totally absorbing boundary condition for terminating each round of diffusion, we assume that the particle is instantaneously shifted a distance $\epsilon$ from the interface following each absorption event.

Suppose that the particle starts at $X(0)=x_0$ and the initial state of the gate is $\sigma(0)=k_0$. Let $ \rho^{(\epsilon)}_{k|k_0}(x,t|x_0)$ denote the full propagator for fixed $\epsilon>0$ and initial position $x_0\neq 0$. Similarly, let $ \rho^{(\epsilon)}_{k|1}(x,t|0)$ denote the full propagator when starting at the origin with the gate open.
The first renewal equation takes the form
\numparts
 \begin{eqnarray}
\fl  \rho^{(\epsilon)}_{k|k_0}(x,t|x_0)&=\Pi_{kk_0}(t)p(x,t|x_0)\Theta(x) +\int_0^t \rho^{(\epsilon)}_{k|-1}(x,t-\tau |\epsilon)\Pi_{-1,k_0}(\tau)f(x_0,\tau)d\tau,
 \nonumber \\
 \label{renintera}
 \fl &  \ +\frac{1}{2} \int_0^t  \bigg [ \rho^{(\epsilon)}_{k|1}(x,t-\tau |\epsilon) +\rho^{(\epsilon)}_{k|1}(x,t-\tau |-\epsilon) \bigg ]\Pi_{1,k_0}(\tau)f(x_0,\tau)d\tau,
 \end{eqnarray}
 for $x_0>0$, and
 \begin{eqnarray}
 \fl \rho^{(\epsilon)}_{k|k_0}(x,t|x_0)&=\Pi_{kk_0}(t)p(-x,t|-x_0)\Theta(-x) \nonumber \\
 \fl &\quad + \int_0^t \rho^{(\epsilon)}_{k|-1}(x,t-\tau |-\epsilon)\Pi_{-1,k_0}(\tau)f(-x_0,\tau)d\tau
  \label{reninterb}\\
\fl   &  \quad +\frac{1}{2} \int_0^t   \bigg [ \rho^{(\epsilon)}_{k|1}(x,t-\tau |\epsilon) +\rho^{(\epsilon)}_{k|1}(x,t-\tau |-\epsilon) \bigg ]\Pi_{1,k_0}(\tau)f(-x_0,\tau)d\tau ,  \nonumber  \end{eqnarray}
   \endnumparts
  for $x_0<0$, where $\Theta(x)$ is a Heaviside function.
   Here $p(x,t|x_0)$ is the propagator for Brownian motion on the half-line $[0,\infty)$ with a totally absorbing boundary at $x=0$:
   \begin{eqnarray}
   \label{pq}
 \fl   \frac{\partial p}{\partial t}=D \frac{\partial^2 p}{\partial x^2},\quad x>0, \quad p(0^+,t|x_0)=0,\quad p(x,0|x_0)=\delta(x-x_0),\quad x,x_0>0.   
 \end{eqnarray}
Moreover, $f(x_0,t)=D\partial_xp(0^+,t|x_0)$ with $x_0>0$ is the corresponding FPT density for absorption. The first term on the right-hand side of equation (\ref{renintera}) represents all sample trajectories starting at $x_0>0$ and ending at $x>0$ that have never reached the interface up to time $t$. If $x<0$ then this contribution vanishes. The corresponding integral terms represent all trajectories that are first absorbed at time $\tau$ and then either restart on the right-hand side at $x=\epsilon$ when the gate is closed or restart at $x=\pm \epsilon$ with equal probability when the gate is open. Similarly, the first term on the right-hand side of equation (\ref{reninterb}) represents all sample trajectories starting at $x_0<0$ and ending at $x<0$ that have never reached the interface up to time $t$. The corresponding integral terms now represent all trajectories that are first absorbed at time $\tau$ and then either restart on the left-hand side at $x=-\epsilon$ when the gate is closed or restart at $x=\pm \epsilon$ with equal probability when the gate is open.
\subsection*{Some remarks}

\noindent (a) As we show below, the renewal equation given by (\ref{renintera}) and (\ref{reninterb}) are equivalent to the forward Kolmogorov equation (\ref{1Da})--(\ref{1Dd}) in the limit $\epsilon\rightarrow 0$. However, the former provides a more general probabilistic framework for modelling BM through a stochastically-gated interface. For example, keeping $\epsilon>0$ fixed represents a form of boundary-induced resetting \cite{Bressloff25b}. That is, whenever the particle encounters the interface it immediately resets to a point that is a distance $\epsilon$ from the interface. More generally, we could take the reset positions on either side of the interface to be different. 
\medskip

\noindent (b) An important general feature of the renewal equation is that its basic structure still holds under various modifications of the diffusion process prior to each absorption event. The only difference is in the construction of the densities $p(x,t|x_0)$ and $f(x_0,t)$. For example, the particle could be confined to a finite domain $[-L,L]$, say, with reflecting or sticky boundaries at $x=\pm L$ and an interface at $x=0$. Alternatively, the particle could reset its position at a random sequence of times generated by a Poisson process along analogous lines to snapping out BM \cite{Bressloff22}. This bulk resetting is distinct from boundary-induced resetting. We discuss this further in section 4. Finally, the detection of the gated interface during each round of BM could be generalised by taking each side of the barrier to be partially rather than totally absorbing.
\medskip

\subsection{Solution in Laplace space for finite $\epsilon>0$}

It is convenient to define the composite densities
 \begin{eqnarray}
  \label{PiG2}
 \fl p_{k|k_0}(x,t|x_0)&=\Pi_{kk_0}(t)p(x,t|x_0) ,\quad  f_{k|k_0}(x_0,t)=\Pi_{kk_0}(t)f(x_0,t) \mbox{ for } x,x_0>0,
\end{eqnarray}
 with $\Pi_{kk_0}(t)$ given by equation (\ref{Pi}). Note that 
 \begin{equation}
 \sum_{k=\pm 1} p_{k|k_0}(x,t|x_0)=p(x,t|x_0) \mbox{ and } \sum_{k=\pm 1} f_{k|k_0}(x_0,t)=f(x_0,t),
 \end{equation}
since $\sum_{k=\pm 1}\Pi_{kk_0}(t)=1$. 
 Laplace transforming equations (\ref{renintera}) and (\ref{reninterb}) with respect to time $t$ and using the convolution theorem then gives
 \numparts
 \begin{eqnarray}
\fl  \widetilde{\rho}^{(\epsilon)}_{k|k_0}(x,s|x_0) &= \p_{k|k_0}(x,s|x_0)\Theta(x)+ \Q^{(\epsilon)}_{k }(x,s) \f_{-1|k_0}(x_0,s) \nonumber  \\
 \label{renewal2a}
 \fl & \quad +  \PP^{(\epsilon)}_k(x,s)\f_{1|k_0}(x_0,s)  ,\ x_0>0 \\
  \fl \widetilde{\rho}^{(\epsilon)}_{k|k_0}(x,s|x_0) &=   \p_{k|k_0}(-x,s|-x_0)\Theta(-x)+ \widetilde{R}^{(\epsilon)}_{k }(x,s) \f_{-1|k_0}(-x_0,s)\nonumber  \\
 \label{renewal2b}
  \fl  &\quad  +\PP^{(\epsilon)}_k(x,s) \f_{1|k_0}(-x_0,s) ,\ x_0<0,
 \end{eqnarray}
 \endnumparts
 where we have set
 \numparts
  \begin{eqnarray}
\PP^{(\epsilon)}_k(x,s)&=\frac{1}{2} \bigg [\wrho^{(\epsilon)}_{k|1}(x,s|\epsilon) +\wrho^{(\epsilon)}_{k|1}(x,s|-\epsilon) \bigg ],\\
 \Q^{(\epsilon)}_{k }(x,s)&=\widetilde{\rho}^{(\epsilon)}_{k|- 1}(x,s|\epsilon),\ \widetilde{R}^{(\epsilon)}_{k }(x,s)=\widetilde{\rho}^{(\epsilon)}_{k|- 1}(x,s|-\epsilon) .
\end{eqnarray}
\endnumparts
Moreover,
 \begin{eqnarray}
\p_{k |k_0 }(x,s|x_0) 
&=\int_0^{\infty}\e^{-st}\Pi_{kk_0}(t) p(x,t|x_0)dt\nonumber \\
&= \int_0^{\infty}\e^{-st}\bigg [\sigma_k+ (\delta_{k,k_0}-\sigma_k)\e^{-(\alpha+\beta) t }\bigg ]   p(x,t|x_0)dt \nonumber\\
&=\sigma_k\p(x,s|x_0)+kk_0\sigma_{-k_0}\p(x,s+\alpha+\beta |x_0) ,
\label{pk}
\end{eqnarray}
and similarly for $\f_{k |k_0 }(x_0,s) $. Laplace transforming (\ref{pq}) shows that 
\begin{equation}
\p(x,s|x_0)=\G(x,s|x_0),\quad \f(x_0,s)=\e^{-\sqrt{s/D}x_0},
\end{equation}
where 
$\G(x,s|x_0) $ is the Dirichlet Green's function (\ref{DG}). 
 
 Next, setting $(x_0,k_0)=(\epsilon,- 1)$ in equation (\ref{renewal2a}) and $(x_0,k_0)=(-\epsilon,- 1)$ in equation (\ref{renewal2b}) yields  
\numparts
\begin{eqnarray}
\fl \Q^{(\epsilon)}_k(x,s)&=\p_{k|-1 }(x,s|\epsilon)\Theta(x)+  \Q^{(\epsilon)}_{k}(x,s) \f_{-1|-1}(\epsilon,s) +  \widetilde{P}^{(\epsilon)}_{k}(x,s)  \f_{1|-1}(\epsilon,s),\\
\fl \widetilde{R}^{(\epsilon)}_k(x,s)&=\p_{k|-1 }(-x,s|\epsilon)\Theta(-x)+ \widetilde{R}^{(\epsilon)}_{k}(x,s) \f_{-1|-1}(\epsilon,s) +     \widetilde{P}^{(\epsilon)}_{k}(x,s)\f_{1|-1}(\epsilon,s).\nonumber \\
\fl
\end{eqnarray}
\endnumparts
Rearranging these equations gives 
\numparts
\begin{eqnarray}
\label{PaQaa}
 \fl &Q^{(\epsilon)}_{k}(x,s)=\frac{\p_{k|- 1}(x,s|\epsilon)\Theta(x)}{1-\f_{-1|-1}(\epsilon,s)}+ \frac{\f_{1|-1}(\epsilon,s)}{1-\f_{-1|-1}(\epsilon,s)}\widetilde{P}^{(\epsilon)}_{k}(x,s),\\
\label{PaQab}
\fl &\widetilde{R}^{(\epsilon)}_{k}(x,s)=\frac{\p_{k|- 1}(-x,s|\epsilon)\Theta(-x)}{1-\f_{-1|-1}(\epsilon,s)}+\frac{\f_{1|-1}(\epsilon,s)}{1-\f_{-1|-1}(\epsilon,s)}\widetilde{P}^{(\epsilon)}_k(x,s).
 \end{eqnarray}
 \endnumparts
Similarly, setting $(x_0,k_0)=(\epsilon,1)$ in equation (\ref{renewal2a}), $(x_0,k_0)=(-\epsilon,1)$ in equation (\ref{renewal2b}) and adding the results gives
\begin{eqnarray}
\fl \PP^{(\epsilon)}_k(x,s)&=\frac{1}{2} \p_{k|1}(|x|,s|\epsilon)+\frac{1}{2} \bigg [\Q^{(\epsilon)}_{k }(x,s) +\widetilde{R}^{(\epsilon)}_{k }(x,s) \bigg ]\f_{-1|1}(\epsilon,s)  +  \PP^{(\epsilon)}_k(x,s)\f_{1|1}(\epsilon,s) ,\nonumber \\
\fl
\end{eqnarray}
Combining with equations (\ref{PaQaa}) and (\ref{PaQab}) and rearranging, we have
\begin{eqnarray}
\fl \PP^{(\epsilon)}_k(x,s)&=\frac{1}{2\Lambda(\epsilon,s)} \bigg [\p_{k|1}(|x|,s|\epsilon)+p_{k|-1}(|x|,s|\epsilon) \frac{\f_{-1|1}(\epsilon,s)}{1-\f_{-1|-1}(\epsilon,s)} \bigg ],
  \label{PaQac}
\end{eqnarray}
with
\begin{equation}
\label{Lam}
\Lambda(\epsilon,s)= 1-  \f_{1|1}(\epsilon,s) -\frac{\f_{1|-1}(\epsilon,s)\f_{-1|1}(\epsilon,s)}{1-\f_{-1|-1}(\epsilon,s)}  .
\end{equation}
In summary, the explicit solution of the Laplace transformed renewal equations is given by equations (\ref{renewal2a}), (\ref{renewal2b}), (\ref{PaQaa}), (\ref{PaQab}) and (\ref{PaQac}).

\subsection{Equivalence to the Kolmogorov equation in the limit $\epsilon\rightarrow 0$}

In order to compare with the solution of the forward Kolmogorov equation derived in section 2 we take the limit $\epsilon\rightarrow 0$ in equations (\ref{renintera}), (\ref{reninterb}), (\ref{PaQaa}), (\ref{PaQab}) and (\ref{PaQac}). Using the expansions
\numparts
\begin{eqnarray}
\fl \f_{k|k_0}(\epsilon,s)&=\sigma_k\e^{-\sqrt{s/D}\epsilon }+kk_0\sigma_{-k_0}\e^{-\sqrt{[s+\alpha+\beta] /D}\epsilon}\nonumber\\
\fl &=\delta_{k,k_0}-\bigg [\sqrt{s/D}\sigma_k +kk_0\sigma_{-k_0}\sqrt{[s+\alpha+\beta] /D}\bigg ]\epsilon+O(\epsilon^2),\\
\fl \p_{k|k_0 }(x,s|\epsilon)&=\bigg [\sigma_k\e^{-\sqrt{s/D}x }+kk_0\sigma_{-k_0}\e^{-\sqrt{[s+\alpha+\beta] /D}x}\bigg ]\frac{\epsilon}{D}+O(\epsilon^2).
\end{eqnarray}
\endnumparts
we find that
\numparts
\begin{eqnarray}
   \Q^{(0)}_{k}(x,s)&= \Theta(x) \frac{\sigma_k \e^{-\sqrt{s/D} x} -k\sigma_{1} \e^{- \sqrt{[s+\alpha+\beta]/D} x}}{\sigma_{-1}\sqrt{sD}+\sigma_1\sqrt{[s+\alpha+\beta]D}},\nonumber\\
\label{limPaQaa}
  &\quad +  \frac{\sigma_{1}  \sqrt{s+\alpha+\beta}-\sigma_1 \sqrt{s} }{\sigma_{-1}\sqrt{s}+\sigma_1 \sqrt{s+\alpha+\beta}} \widetilde{P}^{(0)}_k(x,s)\\
    \widetilde{R}^{(0)}_{k}(x,s)&= \Theta(-x) \frac{\sigma_k \e^{-\sqrt{s/D} |x|} -k\sigma_{1} \e^{- \sqrt{[s+\alpha+\beta]/D} |x|}}{\sigma_{-1}\sqrt{sD}+\sigma_1\sqrt{[s+\alpha+\beta]D}}\nonumber \\
 &\quad +  \frac{\sigma_{1}  \sqrt{s+\alpha+\beta}-\sigma_1 \sqrt{s} }{\sigma_{-1}\sqrt{s}+\sigma_1 \sqrt{s+\alpha+\beta}} \widetilde{P}^{(0)}_k(x,s).
 \label{limPaQab}
\end{eqnarray}
\endnumparts
with
\begin{equation}
\widetilde{P}^{(0)}_k(x,s) =\sigma_k\frac{\e^{-\sqrt{s/D}|x|}}{2\sqrt{sD}}+k\sigma_{-1} \frac{\e^{-\sqrt{(s+\alpha+\beta)/D}|x|}}{2\sqrt{(s+\alpha+\beta)D}}.
\label{Pk}
\end{equation}
Substituting the expressions (\ref{limPaQaa}) and (\ref{limPaQab}) back into equations (\ref{renewal2a}) for $x_0>0$, after taking the limit $\epsilon\rightarrow 0$, yields
 \begin{eqnarray}
  \label{renewal3m}
\fl  & \widetilde{\rho}_{k|k_0}(x,s|x_0)\\
\fl  &= Q_{k}^{(0)}(x,s) \f_{-1|k_0}(x_0,s) +\widetilde{P}^{(0)}_k(x,s) \f_{1|k_0}(x_0,s) \nonumber \\
 \fl    &= \widetilde{P}^{(0)}_k(x,s) \bigg (\frac{\sigma_{1}  \sqrt{s+\alpha+\beta}-\sigma_1 \sqrt{s} }{\sigma_{-1}\sqrt{s}+\sigma_1 \sqrt{s+\alpha+\beta}}\bigg [\sigma_{-1}\e^{-\sqrt{s/D}x_0 }-k_0\sigma_{-k_0}\e^{-\sqrt{[s+\alpha+\beta] /D}x_0}\bigg]\nonumber \\
  \fl &\hspace{3cm} + \bigg [\sigma_{1}\e^{-\sqrt{s/D}x_0 }+k_0\sigma_{-k_0}\e^{-\sqrt{[s+\alpha+\beta] /D}x_0}\bigg]\bigg )\nonumber \\
  \fl &=\widetilde{P}^{(0)}_k(x,s) \frac{\sigma_{1}\sqrt{s+\alpha+\beta}\e^{-\sqrt{s/D}x_0 }+k_0\sigma_{-k_0}\sqrt{s}\e^{-\sqrt{[s+\alpha+\beta] /D}x_0} }{\sigma_{-1}\sqrt{s}+\sigma_1 \sqrt{s+\alpha+\beta}}  .\nonumber
 \end{eqnarray}
 for $x<0$. Similarly, for $x>0$ we have
\begin{eqnarray}
  \label{renewal3p}
  \fl &\widetilde{\rho}_{k|k_0}(x,s|x_0) \\
  \fl &=\p_{k|k_0}(x,s|x_0)\Theta(x)+  \Q_{k}^{(0)}(x,s) \f_{-1|k_0}(x_0,s) +\widetilde{P}^{(0)}_k(x,s) \f_{1|k_0}(x_0,s) \nonumber \\
  \fl   &= \sigma_{k} \G(x,s|x_0)+k k_0\sigma_{-k_0}\G(x,s+\alpha+\beta|x_0)\nonumber \\
  \fl &\quad +\widetilde{P}^{(0)}_k(x,s) \frac{\sigma_{1}\sqrt{s+\alpha+\beta}\e^{-\sqrt{s/D}x_0 }+k_0\sigma_{-k_0}\sqrt{s}\e^{-\sqrt{[s+\alpha+\beta] /D}x_0} }{\sigma_{-1}\sqrt{s}+\sigma_1 \sqrt{s+\alpha+\beta}}  .\nonumber \\
  \fl &\quad +\frac{\sigma_k \e^{-\sqrt{s/D} x} -k\sigma_{1} \e^{- \sqrt{s+\alpha+\beta]/D} x}}{\sigma_{-1}\sqrt{sD}+\sigma_1\sqrt{[s+\alpha+\beta]D}}\bigg [\sigma_{-1}\e^{-\sqrt{s/D}x_0 }-k_0\sigma_{-k_0}\e^{-\sqrt{[s+\alpha+\beta] /D}x_0}\bigg ]\nonumber 
 \end{eqnarray}
  After some algebra, it can be checked that equations (\ref{renewal3m}) and (\ref{renewal3p}) with  $\widetilde{P}^{(0)}_k(x,s)$ given by equation (\ref{Pk}) are consistent with equations (\ref{sola}) and (\ref{solb}). The corresponding solutions for $x_0<0$ can be obtained by combining equations (\ref{PaQaa}) and (\ref{PaQab}) with (\ref{renewal2b}).

Finally, it is interesting to note that the inverse Laplace transform of equation (\ref{Pk}) is
\begin{equation}
 \rho_{k|1}(x,t |0)=P^{(0)}_k(x,t)\equiv \Pi_{k1}(t)P(x,t|0),
\end{equation}
where $P(x,t|0)= \e^{-x^2/4Dt}/\sqrt{4\pi Dt}.$ is the propagator for BM in $\R$. That is, if the particle starts at the origin and the gate is open then the solution is an even function of $x$, $\rho_{k|1}(x,t |0)=\rho_{k|1}(-x,t |0)$ for all $x\geq 0$ and $k=\pm 1$. It follows that the propagator is continuous at $x=0$. In other words, the marginal density for particle position is independent of the gate dynamics.

\section{NESS with stochastic resetting}

Now suppose that BM in the positive half line is supplemented by the reset condition $X(t) \rightarrow \xi \in   [0^+,\infty)$ at a random sequence of times generated by a Poisson process with constant rate $r$. This particular problem was originally studied in Refs.  \cite{Evans11a,Evans11b}. Equation (\ref{pq}) becomes
\numparts
\begin{eqnarray}
\label{r1Da}
 &\frac{\partial p}{\partial t}=D\frac{\partial^2p}{\partial x^2} -rp+r S(x_0,t)\delta(x-\xi),\quad x>0,\\
& p(0,t|x_0) =0,\quad
 p(x,0|x_0) = \delta(x - x_0).
\label{r1Db}
\end{eqnarray}
\endnumparts
We have introduced the marginal distribution
\begin{equation}
\label{r1DQ}
S(x_0,t)=\int_0^{\infty} p(x,t|x_0)dx,
\end{equation}
which is the survival probability that the particle hasn't been absorbed at $x=0$ in the time interval $[0,t]$, having started at $x_0$. Laplace transforming equations (\ref{1Da}) and (\ref{1Db}) gives
\numparts
\begin{eqnarray}
\label{r1DLTa}
\fl   &D\frac{\partial^2\p(x,s|x_0)}{\partial x^2} -(r+s)p(x,s|x_0)=-[\delta(x-x_0)+r \S(x_0,s)\delta(x-\xi)],\ x>0,\\
\fl  &\p(0,s|x_0) =0.
\label{r1DLTb}
\end{eqnarray}
\endnumparts
In terms of the Dirichlet Green's function (\ref{DG}), we have
\begin{eqnarray}
\label{rp1D}
\fl    \p(x, s|x_0) =  \G(x,r+s|x_0)+ r\S(x_0,s)\G(x,r+s|\xi), \quad 0<x<\infty.
\end{eqnarray}
Next, Laplace transforming equation (\ref{r1DQ}) and using (\ref{rp1D}) shows that
 \begin{eqnarray}
  \S(x_0,s)&=\int_0^{\infty}\p(x,s|x_0)dx\nonumber \\
  &=\int_0^{\infty} \G(x,r+s|x_0)dx+  r\S(x_0,s)\int_0^{\infty} \G(x,r+s|\xi)dx\nonumber \\
&=\S_0(x_0,r+s)+r\S(x_0,s)\S_0(\xi,r+s),
\label{QQr}
 \end{eqnarray}
 where $\S_0$ is the Laplace transform of the survival probability without resetting:
 \begin{equation}
 \label{QQ0}
 \S_0(x_0,s)=\frac{1-\e^{-\sqrt{s/D}x_0}}{s}.
 \end{equation}
  Rearranging equation (\ref{QQr}) thus determines the survival probability with resetting in terms of the corresponding probability without resetting:
 \begin{equation}
 \label{Qr}
 \S(x_0,s)=\frac{\S_0(x_0,r+s)}{1-r\S_0(\xi,r+s)}.
 \end{equation}
Finally, since the FPT density $f(x_0,t)=-\partial_tS(x_0,t)$ so that $\f(x_0,s)= 1-s\S(x_0,s)$, we have
\begin{equation}
\label{fr}
\f(x_0,s)= \frac{r\e^{-\sqrt{(r+s)/D}\xi}+s\e^{-\sqrt{(r+s)/D}x_0}}{s+r\e^{-\sqrt{(r+s)/D}\xi}}.
\end{equation}

\begin{figure}[t!]
  \centering
  \includegraphics[width=10cm]{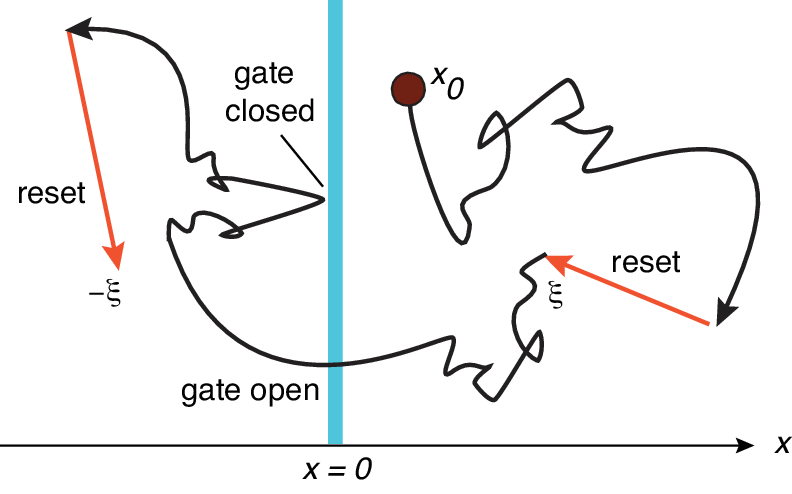}
  \caption{BM through a stochastically gated interface with stochastic resetting to $\pm \xi$. (The dynamics is extended into two dimensions for ease of visualization.) The particle starts on the right-hand side of the interface and undergoes one reset to $\xi$ before passing through an open gate to the left-hand side. Whilst in this domain the particle reflects off a closed gate and then resets to $-\xi$ etc. Resetting events that cross the membrane are forbidden.}
  \label{fig4}
\end{figure}

For simplicity, we will assume BM in the negative half line resets to the mirror image $-\xi$ so that the reflection symmetry assumed in the renewal equations (\ref{renintera}) and (\ref{reninterb}) is preserved, see Fig. \ref{fig4}. That is, we have a space-dependent resetting protocol in which $X(t) \geq 0^+$ ($X(t) \leq 0^-$) can only reset to $x=\xi$ ($x=-\xi$). In particular, the particle cannot cross the interface by resetting. It follows that the renewal equations (\ref{renewal2a}) and (\ref{renewal2b}) still hold except that $\p(x,s|x_0)$ and $\f(x_0,s)$ are given by equations (\ref{rp1D}) and (\ref{fr}), respectively. 
One of the characteristic features of non-absorbing diffusion processes with stochastic resetting is that there exists a nonequilibrium stationary state (NESS), which is maintained by non-zero probability fluxes \cite{Evans20}. In the case of the resetting protocol shown in Fig. \ref{fig4},  the points $x=\pm \xi$ act as probability sources, whereas all positions $x\neq \pm \xi$ are potential probability sinks. Although each round of BM on the half-line is terminated by absorption at the interface, the stochastic process is immediately restarted so that BM across the stochastically gated interface is not terminated and supports an NESS. We will derive the latter using the renewal equations (\ref{renewal2a}) and \ref{renewal2b}) together with the observation that BM on the half-line with resetting and an absorbing boundary at $x=0$ does not have a non-trivial NESS, that is,
 \begin{equation}
 \lim_{t\rightarrow \infty} p(x,t|x_0)=\lim_{s\rightarrow 0}s\p(x,s|x_0)=0.
 \end{equation}
 
 \begin{figure}[b!]
  \centering
  \includegraphics[width=10cm]{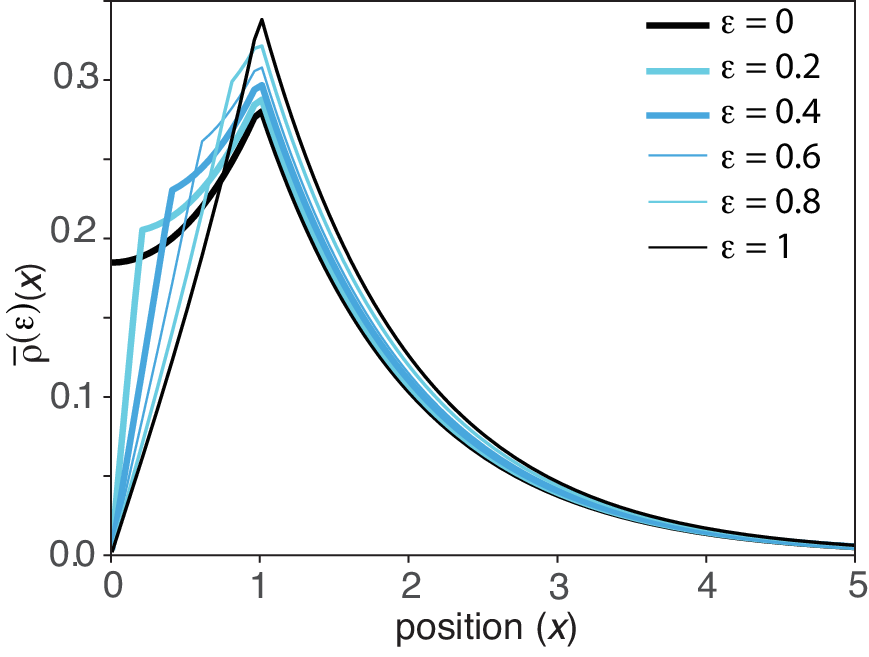}
  \caption{NESS $\overline{\rho}^{(\epsilon)}(x))$ for 1D BM with a stochastically-gated interface at $x=0$, bulk resetting to $\pm \xi$ at a rate $r$ (see Fig.\ref{fig4}), and boundary-induced resetting with a shift of size $\epsilon$. Since $\overline{\rho}^{(\epsilon)}(x)$ is an even function of $x$ we show plots for various values of $\epsilon$ with $x\geq 0$. Other parameters are $D=\xi=1$, $\alpha=\beta=1$ and $ r=1$.}
  \label{fig5}
\end{figure}

\begin{figure}[t!]
  \centering
  \includegraphics[width=10cm]{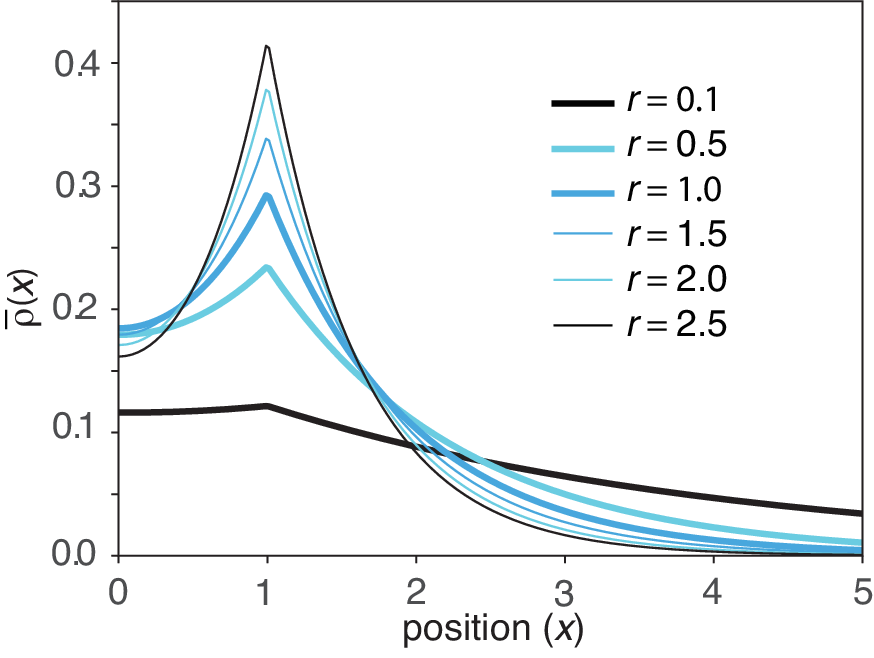}
  \caption{Same as Fig. \ref{fig5} for $\epsilon=0$ and different values of the resetting rate $r$.}
  \label{fig6}
\end{figure}

\begin{figure}[b!]
  \centering
  \includegraphics[width=10cm]{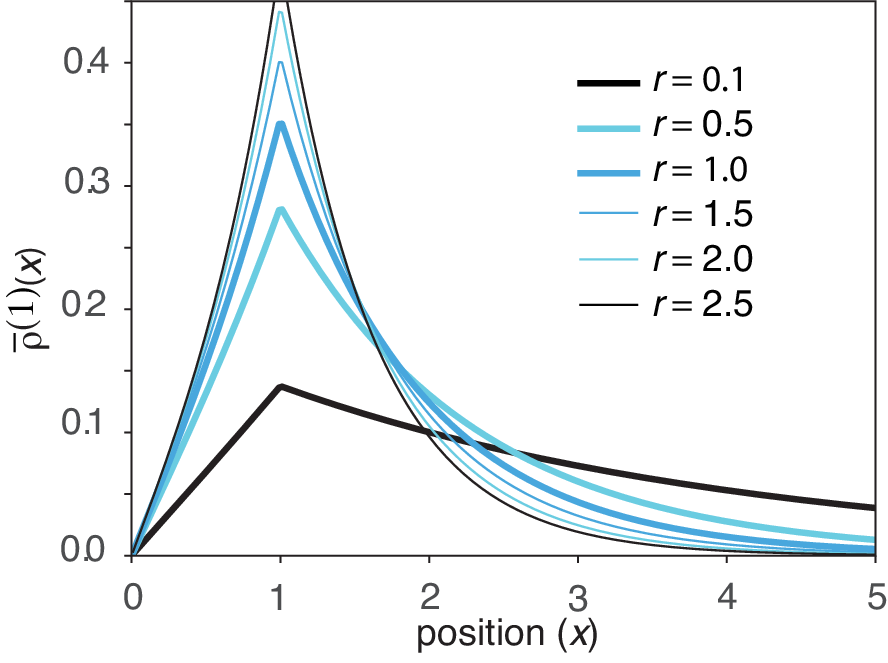}
  \caption{Same as Fig. \ref{fig5} for $\epsilon=1$ and different values of the resetting rate $r$.}
  \label{fig7}
\end{figure}

 Multiplying both sides of (\ref{renewal2a}) by $s$ and taking the limit $s\rightarrow 0$ gives
 \begin{eqnarray}
 \label{ss1}
 \fl  \overline{\rho}^{(\epsilon)}_{k}(x) &\equiv \lim_{s\rightarrow 0} s \widetilde{\rho}^{(\epsilon)}_{k|k_0}(x,s|x_0) =  \overline Q^{(\epsilon)}_{k}(x)  \f_{-1|k_0}(x_0,0) +\overline P^{(\epsilon)}_{k}(x)\f_{1|k_0}(x_0,0)  \label{reseta}
  \end{eqnarray}
  where
  \begin{equation}
  \label{Qstar}
 \fl  \overline Q^{(\epsilon)}_{k}(x)= \frac{\f_{1|-1}(\epsilon,0)}{1-\f_{-1|-1}(\epsilon,0)}\overline{P}^{(\epsilon)}_{k}(x),\quad \overline{P}^{(\epsilon)}_{k}(x)=\lim_{s\rightarrow 0} s \PP^{(\epsilon)}_k(x,s)
  \end{equation}
  and $ \PP^{(\epsilon)}_k(x,s)$ is given by equation (\ref{PaQac}). Using equation (\ref{fr}), we have
\begin{eqnarray}
\fl \f_{k|k_0}(\epsilon,0)&=\sigma_k+kk_0\sigma_{-k_0}\bigg [ \frac{r\e^{-\sqrt{(r+\alpha+\beta)/D}\xi}+(\alpha+\beta)\e^{-\sqrt{(r+\alpha+\beta)/D}\epsilon}}{\alpha+\beta+r\e^{-\sqrt{(r+\alpha+\beta)/D}\xi}}\bigg ]\nonumber \\
\fl & \equiv \sigma_k+kk_0\sigma_{-k_0} \Gamma_r(\epsilon).
\end{eqnarray}
It follows that $\overline Q^{(\epsilon)}_{k}(x)=\overline{P}^{(\epsilon)}_{k}(x)$ and $\f_{1|k_0}(x_0,0) +\f_{-1|k_0}(x_0,0) =1$. Hence,
\begin{eqnarray}
\overline{\rho}^{(\epsilon)}_{k}(x) =\overline{P}^{(\epsilon)}_{k}(x),
\end{eqnarray}
which is independent of $x_0$ and $k_0$ as expected.
Finally, in order to take the limit $s\rightarrow 0$ in equation (\ref{PaQac}) we note that
\begin{eqnarray}
\Lambda(\epsilon,0)&=1-\sigma_1-\sigma_{-1}\Gamma_r(\epsilon)-\frac{\sigma_1\sigma_{-1}(1-\Gamma_r(\epsilon))^2}{1-\sigma_{-1}-\sigma_1 \Gamma_r(\epsilon)} 
 =0.
\end{eqnarray}
Therefore, we obtain the following formula for the NESS
\begin{eqnarray}
\fl \overline{\rho}^{(\epsilon)}_{k}(x)&=\frac{1}{2\partial_s\Lambda(\epsilon,0)} \bigg [\p_{k|1}(|x|,0|\epsilon)+p_{k|-1}(|x|,0|\epsilon) \frac{\f_{-1|1}(\epsilon,0)}{1-\f_{-1|-1}(\epsilon,0)} \bigg ].
\end{eqnarray}

In Figs. \ref{fig5}--\ref{fig7} we show example plots of the NESS $\overline{\rho}^{(\epsilon)}(x)=\sum_{k=\pm 1}\overline{\rho}_k^{(\epsilon)}(x) $ as a function of $x$. Since the NESS is an even function of $x$ we show plots for $x\geq 0$. We first consider the dependence of $\overline{\rho}^{(\epsilon)}(x)$ on the shift $\epsilon$ for fixed resetting rate $r$, see Fig. \ref{fig5}. It can be seen that for $\epsilon$ the flux is discontinuous at $x=0$ and the two reset points $x=\epsilon$ and $x=\xi$ with $\overline{\rho}^{(\epsilon)}(0)=0$. On the other hand, $\overline{\rho}(x)=\lim_{\epsilon \rightarrow 0}\overline{\rho}^{(\epsilon)}(x)$ is smooth function of $x$ at the origin and only has a cusp at $x=\xi$. In Figs. \ref{fig6} and  \ref{fig7} we plot $\overline{\rho}(x)$ and $\overline{\rho}^{(1)}(x)$, respectively, for different values of $r$.

\section{Higher-dimensional interface in $\R^d$}

\begin{figure}[h!]
  \centering
  \includegraphics[width=6cm]{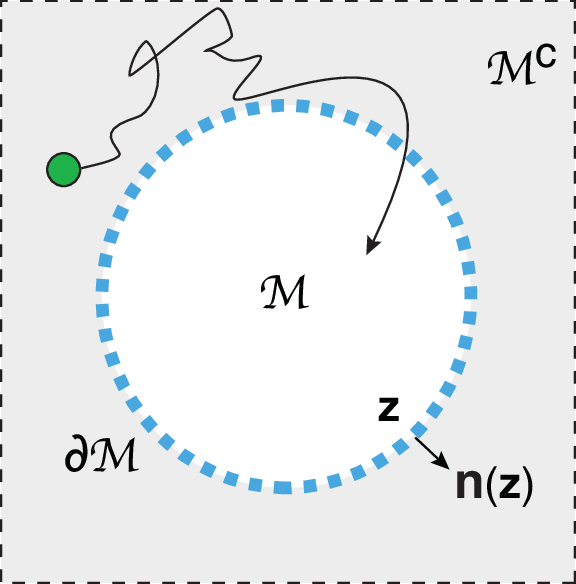}
  \caption{BM in $\R^d$ with a stochastically-gated interface $\partial \calM\subset \R^{d-1}$.}
  \label{fig8}
\end{figure}

Consider a Brownian particle diffusing in $\R^d$, which is partitioned into two domains by a stochastically gated interface $\partial \calM\subset R^{d-1}$, where $\calM$ is a bounded simply connected set, see Fig. \ref{fig8}. Let $\n(\y)$ denote the outward unit normal at a point $\y \in \partial \calM$ and define the outer and inner surfaces of $\partial \calM$ according to
\begin{equation}
\partial \calM^{\pm} =\lim_{\epsilon \rightarrow 0} \{\y \pm \epsilon\n(\y),\quad \y \in \partial \calM\}.
\end{equation}
Let $\rho^{(\epsilon)}_{k|k_0}(\x,t|\x_0)$, $\x,\x_0\in \calM\cup \calM^c$, denote the joint propagator density of the particle and gate under the initial condition $\X(0)=\x_0,\sigma(0)=k_0$. The higher-dimensional version of the forward Kolmogorov equation (\ref{1Da})--(\ref{1Dd}) takes the form
\numparts
\begin{eqnarray}
\fl &\frac{\partial \rho_{1|k_0}(\x,t|\x_0)}{\partial t}=D\nabla^2\rho_{1|k_0}(\x,t|\x_0) -\alpha \rho_{1|k_0}(\x,t|\x_0)+\beta \rho_{-1|k_0}(\x,t|\x_0),\\
\fl & \frac{\partial \rho_{-1|k_0}(\x,t|\x_0)}{\partial t}=D\nabla^2\rho_{-1|k_0}(\x,t|\x_0) +\alpha \rho_{1|k_0}(\x,t|\x_0)-\beta \rho_{-1|k_0}(\x,t|\x_0),\\
\fl & {\bm \nabla} \rho_{-1|k_0}(\y^+,t|\x_0)=0,\quad {\bm \nabla} \rho_{-1|k_0}(\y^-,t|\x_0)=0,\\
\fl &\rho_{1|k_0}(\y^+,t|\x_0)=  \rho_{-1|k_0}(\y^-,t|\x_0) ,\quad {\bm \nabla} \rho_{1|k_0}(\y^+,t|\x_0)={\bm \nabla} \rho_{1|k_0}(\y^-,t|\x_0),
\end{eqnarray}
\endnumparts
with $\x,\x_0\in \calM\cup \calM^c$ and $y^{\pm} \in \partial \calM^{\pm}$. In this section we construct the higher-dimensional version of the renewal equations (\ref{renintera}) and (\ref{reninterb}), and indicate some of the mathematical challenges arising from the fact that the interface is now spatially extended.

\subsection{Renewal equation}

Suppose that the particle begins at a point $\x_0\in  \calM\cup {\calM^c}$ and the initial state of the gate is $k_0$. The particle diffuses until reaching a point $\y \in \partial \calM$ where it is instantaneously absorbed. The stochastic process is immediately restarted according to the following rules. If the gate is closed then BM restarts at a point on the same side of the interface, that is, $\x=\y+a\n(\y)$ for $\x_0\in \calM^c$ and $\x=\y-a\n(\y)$ for $\x_0\in \calM$. On the other hand, if the gate is open then the particle restarts at $\x=\y\pm \epsilon \n(\y)$ with an equal probability. The first renewal equation for $\x_0\in \calM^c$  
  \numparts
\begin{eqnarray}
  \fl \rho^{(\epsilon)}_{k|k_0}(\x,t|\x_0)&=\Pi_{kk_0}(t)p(\x,t|\x_0){\bf 1}_{\calM^c}(\x)\nonumber \\
  \fl & + \int_0^t d\tau  \int_{\partial \calM} d\y\, \rho^{(\epsilon)}_{k|-1}(\x,t-\tau |\y+a\n(\y)) \Pi_{-1,k_0}(\tau) J(\y,\tau|\x_0) \nonumber \\
  \fl & +\frac{1}{2} \int_0^t d\tau \int_{\partial \calM} d\y\, \bigg [ \rho^{(\epsilon)}_{k|1}(\x,t-\tau |\y+\epsilon \n(\y)) +\rho^{(\epsilon)}_{k|1}(\x,t-\tau |\y-\epsilon \n(\y)) \bigg ]\nonumber \\
  \fl &\hspace{4cm} \times \Pi_{1,k_0}(\tau) J(\y,\tau|\x_0)   ,
   \label{drenewala}
   \end{eqnarray}
   and the corresponding equation for $\x_0\in \calM$ is
\begin{eqnarray}
  \fl \rho^{(\epsilon)}_{k|k_0}(\x,t|\x_0)&=\Pi_{kk_0}(t)q(\x,t|\x_0){\bf 1}_{\calM}(\x)\nonumber \\
  \fl & + \int_0^t d\tau  \int_{\partial \calM} d\y\, \rho^{(\epsilon)}_{k|-1}(\x,t-\tau |\y-\epsilon \n(\y)) \Pi_{-1,k_0}(\tau) K(\y,\tau|\x_0) \nonumber \\
  \fl & +\frac{1}{2} \int_0^t d\tau \int_{\partial \calM} d\y\, \bigg [ \rho^{(\epsilon)}_{k|1}(\x,t-\tau |\y+\epsilon \n(\y)) +\rho^{(\epsilon)}_{k|1}(\x,t-\tau |\y-\epsilon \n(\y)) \bigg ]\nonumber \\
  \fl &\hspace{4cm} \times \Pi_{1,k_0}(\tau) K(\y,\tau|\x_0)  , 
   \label{drenewalb}
   \end{eqnarray}
 \endnumparts 
wherte ${\bf 1}_{\mathcal A}(\x) =1$ if $\x\in {\mathcal A}$ and is zero otherwise. Here $p(\x,t|\x_0)$ is the solution of the Kolmogorov equation for Brownian motion in $\calM^c$ with a totally absorbing boundary $\partial \calM$:
 \numparts 
\label{pdiffloc}
\begin{eqnarray}
	&\frac{\partial p(\x,t|\x_0)}{\partial t} = D\nabla^2 p(\x,t|\x_0) \mbox{ for } \x,\x_0 \in  {\calM^c},\\
&p(\y,t|\x_0) =0\mbox{ for } \y\in \partial \calM^c
	\end{eqnarray}
	\endnumparts 
with $p(\x,0|\x_0)=\delta(\x-\x_0)$ and $J(\y,t|\x_0)=D{\bm \nabla} p(\y,t|\x_0)\cdot \n(\y)$ for $\y\in \partial \calM$. Similarly,
 \numparts 
\label{qdiffloc}
\begin{eqnarray}
	&\frac{\partial q(\x,t|\x_0)}{\partial t} = D\nabla^2 q(\x,t|\x_0) \mbox{ for } \x,\x_0 \in  {\calM},\\
&q(\y,t|\x_0) =0\mbox{ for } \y\in \partial \calM,
	\end{eqnarray}
	\endnumparts 
with $q(\x,0|\x_0)=\delta(\x-\x_0)$ and $K(\y,t|\x_0)=-D{\bm \nabla} q(\y,t|\x_0)\cdot \n(\y)$. The first term on the right-hand side of equation (\ref{drenewala}) represents all sample trajectories that have never reached the boundary $\partial \calM$. It vanishes if $\x\in \calM$. The corresponding integral terms represent all trajectories that are first absorbed at a point $\y\in \partial \calM$ at time $t-\tau$ and then either restart at $\x=\y+a\n(\y)$ if the gate is closed or restart at $\x=\y\pm a\n(\y)$ with equal probability if the gate is open. We also have to integrate with respect to all starting positions $\y$ at time $t-\tau$. An analogous interpretation holds for the various terms on the right-hand side of equation (\ref{drenewalb}).

Laplace transforming the renewal equations (\ref{drenewala}) and  (\ref{drenewalb}) with respect to time $t$ gives
\numparts
 \begin{eqnarray}
  \fl \widetilde{\rho}_{k|k_0}^{(\epsilon)}(\x,s|\x_0) &= \p_{k|k_0}(\x,s|x_0) {\bf 1}_{\calM^c}(\x)+\int_{\partial \calM}\Q^{(\epsilon)}_k(\x,s|\y)\J_{-1|k_0}(\y,s|\x_0)d\y \nonumber \\
 \label{mrenewala}
  \fl  &\quad +\int_{\partial \calM}\PP^{(\epsilon)}_k(\x,s|\y)\J_{1|k_0}(\y,s|\x_0)d\y ,\quad \x_0\in \calM^c,\\
 \fl \widetilde{\rho}_{k|k_0}^{(\epsilon)}(\x,s|\x_0) &= \p_{k|k_0}(\x,s|x_0) {\bf 1}_{\calM}(\x)+\int_{\partial \calM}\widetilde{R}^{(\epsilon)}_k(\x,s|\y)\K_{-1|k_0}(\y,s|\x_0)d\y \nonumber \\
 \fl  &\quad +\int_{\partial \calM}\PP^{(\epsilon)}_k(\x,s|\y)\K_{1|k_0}(\y,s|\x_0)d\y ,\quad \x_0\in \calM,
  \label{mrenewalb}
  \end{eqnarray}
  \endnumparts
  with
  \numparts
 \begin{eqnarray}
 \label{pk2D}
\fl &\p_{k|k_0}(\x,s|\x_0)= \sigma_k\G_+(\x,s|\x_0)+kk_0\sigma_{-k_0}\G_+(\x,s+\alpha+\beta|\x_0) \ \x,\x_0\in \calM^c,\\
\fl &\q_{k|k_0}(\x,s|\x_0)= \sigma_k\G_-(\x,s|\x_0)+kk_0\sigma_{-k_0}\G_-(\x,s+\alpha+\beta|\x_0) ,\ \x,\x_0\in \calM,
\label{qk2D}
\end{eqnarray}
and (for $\x_0\notin \partial \calM$)
 \begin{eqnarray}
\fl  \widetilde{J}_{k|k_0}(\y,s|\x_0)
 &=D\bigg ( \sigma_k{\bm \nabla} \G_+(\y,s|\x_0)+kk_0\sigma_{-k_0}{\bm \nabla} \G_+(\y,s+\alpha+\beta|\x_0)\bigg )\cdot \n(\y)  ,\\
 \label{K2D}
 \fl  \widetilde{K}_{k|k_0}(\y,s|\x_0)
 &=-D\bigg (\sigma_k {\bm \nabla} \G_-(\y,s|\x_0)+kk_0\sigma_{-k_0}{\bm \nabla} \G_-(\y,s+\alpha+\beta|\x_0)\bigg )\cdot \n(\y) .\nonumber \\
 \fl
  \end{eqnarray}
  \endnumparts
  Here $\G_+(\x,s|\x_0)$ is the Dirichlet Green's function of the modified Helmholtz equation in $\calM^c $:
   \numparts
   \label{G2D}
 \begin{eqnarray}
&D{\bm \nabla}^2 \G(\x,s|\x_0)-s\G(\x,s|\x_0)=-\delta(\x-\x_0),\quad \x \in \calM ^c,\\
& \G(\y,s|\x_0)=0, \quad \y \in \partial \calM.
 \end{eqnarray}
 \endnumparts
 and $\G_-(\x,s|\x_0)$ is the corresponding Dirichlet Green's function in $\calM $.
 We have also introduced the functions
  \numparts
  \begin{eqnarray}
  \fl & \Q_k^{(\epsilon)}(\x,s|\y)=\rho^{(\epsilon)}_{k|-1}(\x,t|\y+\epsilon \n(\y)),\quad  \widetilde{R}_k^{(\epsilon)}(\x,s|\y)=\rho^{(\epsilon)}_{k|-1}(\x,t|\y-\epsilon \n(\y)),\quad \\
  \fl  &\PP_k^{(\epsilon)}(\x,s|\y)=\frac{1}{2}\bigg [\rho^{(\epsilon)}_{k|1}(\x,t|\y+\epsilon \n(\y))+\rho^{(\epsilon)}_{k|1}(\x,t|\y-\epsilon \n(\y))\bigg ].
  \end{eqnarray}
  \endnumparts

Proceeding along similar lines to the 1D case, we can derive self-consistency conditions for the unknown functions $ \Q_k^{(\epsilon)}(\x,s|\y)$, $ \widetilde{R}_k^{(\epsilon)}(\x,s|\y)$ by setting $(\x_0,k_0)=(\z+\epsilon \n(\z),-1)$ in equation (\ref{mrenewala}) and $(\x_0,k_0)=(\z-\epsilon \n(\z),1)$ in equation (\ref{mrenewalb}) with $\z\in \partial \calM$:
\numparts
 \begin{eqnarray}
  \fl  \Q_k^{(\epsilon)}(\x,s|\z) &= \p_{k|-1}(\x,s|\z+\epsilon \n(\z)) {\bf 1}_{\calM^c}(\x)+\int_{\partial \calM}\Q^{(\epsilon)}_k(\x,s|\y)\J_{-1|-1}(\y,s| \z+\epsilon \n(\z))d\y \nonumber \\
 \label{mQa}
  \fl  &\quad +\int_{\partial \calM}\PP^{(\epsilon)}_k(\x,s|\y)\J_{1|-1}(\y,s|\z+\epsilon \n(\z))d\y ,\\
 \fl  \widetilde{R}_k^{(\epsilon)}(\x,s|\z) &= \q_{k|-1}(\x,s| \z-\epsilon \n(\z)) {\bf 1}_{\calM}(\x)+\int_{\partial \calM}\widetilde{R}^{(\epsilon)}_k(\x,s|\y)\K_{-1|-1}(\y,s|\z-\epsilon \n(\z))d\y \nonumber \\
 \fl  &\quad +\int_{\partial \calM}\PP^{(\epsilon)}_k(\x,s|\y)\K_{1|-1}(\y,s|\z-\epsilon \n(\z))d\y .
  \label{mRa}
  \end{eqnarray}
Similarly, setting $(\x_0,k_0)=(\z+\epsilon \n(\z),1)$ in equation (\ref{mrenewala}), $(\x_0,k_0)=(\z-\epsilon \n(\z),1)$ in equation (\ref{mrenewalb}) and adding the results shows that
\begin{eqnarray}
  \fl & \PP_k^{(\epsilon)}(\x,s|\z) =\frac{1}{2}\bigg [ \p_{k|1}(\x,s|\z+\epsilon \n(\z)) {\bf 1}_{\calM^c}(\x) +\q_{k|1}(\x,s| \z-\epsilon \n(\z)) {\bf 1}_{\calM}(\x)\bigg ]\nonumber \\
  \fl &+\int_{\partial \calM}\frac{1}{2}\bigg [\Q^{(\epsilon)}_k(\x,s|\y)\J_{-1|1}(\y,s| \z+\epsilon \n(\z))+\widetilde{R}^{(\epsilon)}_k(\x,s|\y)\K_{-1|1}(\y,s|\z-\epsilon \n(\z))\bigg ]d\y \nonumber \\
 \fl  &\quad +\frac{1}{2}\int_{\partial \calM}\PP^{(\epsilon)}_k(\x,s|\y)\bigg [\J_{1|1}(\y,s|\z+\epsilon \n(\z))+\K_{1|1}(\y,s|\z-\epsilon \n(\z))\bigg ]d\z .  \label{mPa}
  \end{eqnarray}
  \endnumparts
  
\subsection{Spectral decompositions}

 Since the surface $\partial \calM$ is spatially extended when $d>1$, the corresponding renewal equations (\ref{drenewala}) and (\ref{drenewalb}) involve spatial integrals with respect to points $\y \in \partial \calM$. These integrals also appear in the Laplace transformed renewal equations (\ref{mrenewala}) and (\ref{mrenewalb}). The self-consistency equations (\ref{mQa})--(\ref{mPa}) for fixed $\x,s,k$ thus take the form of Fredholm integral equations of the second kind with respect to functions on $\partial \calM$. This can be seen more easily by writing the self-consistency equations in a more compact form that suppresses the dependence on $\x,s,k$:
 \numparts
 \begin{eqnarray}
  &\Q^{(\epsilon)}(\z) = H^{(\epsilon)}_Q(\z)+\L^{(\epsilon)}_{-1,-1}[\Q^{(\epsilon)}](\z) +\L^{(\epsilon)}_{1,-1}[\PP^{(\epsilon)}](\z)  ,
 \label{comQa}\\
  &\widetilde{R}^{(\epsilon)}(\z) =  H^{(\epsilon)}_R(\z)+\overline\L^{(\epsilon)}_{-1,-1}[\widetilde{R}^{(\epsilon)}](\z) +\overline\L^{(\epsilon)}_{1,-1}[\widetilde{P}^{(\epsilon)}](\z),
  \label{comRa}\\
  & \PP^{(\epsilon)}(\z) =H^{(\epsilon)}_P(\z)+ \frac{1}{2}\bigg [\L^{(\epsilon)}_{-1,1}[\Q^{(\epsilon)}](\z) + \overline\L^{(\epsilon)}_{-1,1}[\widetilde{R}^{(\epsilon)}](\z) \bigg ] \nonumber \\
  &\hspace{3cm} +\frac{1}{2}  \bigg [\L^{(\epsilon)}_{1,1}[\PP^{(\epsilon)}](\z) +\overline\L^{(\epsilon)}_{1,1}[\widetilde{P}^{(\epsilon)}](\z)\bigg ] .  \label{comPa}
  \end{eqnarray}
  \endnumparts
  Here
  \numparts
  \begin{eqnarray}
  \fl& H^{(\epsilon)}_Q(\z)= \p_{k|-1}(\x,s|\z+\epsilon \n(\z)) {\bf 1}_{\calM^c}(\x),\\
\fl & H^{(\epsilon)}_R(\z)= \q_{k|-1}(\x,s| \z-\epsilon \n(\z)) {\bf 1}_{\calM}(\x),\\
\fl&   H^{(\epsilon)}_P(\z)=\frac{1}{2}\bigg [ \p_{k|1}(\x,s|\z+\epsilon \n(\z)) {\bf 1}_{\calM^c}(\x) +\q_{k|1}(\x,s| \z-\epsilon \n(\z)) {\bf 1}_{\calM}(\x)\bigg ],
\end{eqnarray}
and we have introduced the linear operators
\begin{eqnarray}
\fl &  \L^{(\epsilon)}_{kk'}[u](\z)=\int_{\partial \calM} u(\y)\J_{k|k'}(\y,s| \z+\epsilon \n(\z))d\y,\\ & \overline{ \L}^{(\epsilon)}_{kk'}[u](\z)=\int_{\partial \calM} u(\y)\K_{k|k'}(\y,s| \z-\epsilon \n(\z))d\y.
    \end{eqnarray}
\endnumparts
We now note that $ \L^{(\epsilon)}_{kk'}$ and $\overline  \L^{(\epsilon)}_{kk'}$ act on the space of functions $L^2(\partial \calM)$ with $\partial \calM$ a compact domain, which implies that they have discrete spectra. Hence, 
one way to formally solve equations (\ref{comQa})--(\ref{comPa} ) is to perform series expansions with respect to the eigenfunctions of one of the linear operators. However, in general, this leads to an infinite-dimensional matrix equation for the coefficients in the expansions of $\Q^{(\epsilon)}(\z) $ etc., so that one would need to numerically truncate the resulting series solution.

Here we use equations (\ref{comQa})--(\ref{comPa} ) to  establish the existence of a solution to the renewal equations (\ref{mrenewala}) and (\ref{mrenewalb}) in the limit $\epsilon \rightarrow 0$. In order to take the latter limit, we introduce the Taylor expansions
\numparts
 \begin{eqnarray}
 \label{Jeps}
  \J(\y,s|\z+\epsilon \n(\z))= \J (\y,s| \z)+ \epsilon   {\bm \nabla }_{\z} \J(\y,s|\z) \cdot \n(\z)+O(\epsilon^2) ,\\
  \K(\y,s|\z-\epsilon \n(\z))= \K (\y,s| \z)- \epsilon{\bm \nabla }_{\z} \K(\y,s|\z) \cdot \n(\z)+O(\epsilon^2) .
   \label{Keps}
 \end{eqnarray}
 \endnumparts
Note that care has to be taken in defining the boundary flux when the initial condition is also on the boundary, see Ref.  \cite{Bressloff23a}. In particular, if $\z\in \partial \calM$ then $\J(\x,s|\z)\equiv D{\bm \nabla} \G(\x,s|\z)\cdot \n(\x)=0$ for all $\x\not \in \partial \calM$ due to the Dirichlet boundary condition $G(\x,s|\z)=0$. On the other hand, $ \J(\y,s| \z)=\overline{\delta}(\y-\z)$ for $\y,\z\in \partial \Omega$ and independently of $s$, where the Dirac delta function is restricted to the surface $\calU$.
It follows that
\begin{eqnarray}
 \L^{(\epsilon)}_{kk'}[u](\z)&=\delta_{k,k'}u(\z)-\epsilon \L_{kk'}[u](\z)+O(\epsilon^2),\\
 \overline \L^{(\epsilon)}_{kk'}[u](\z)&=\delta_{k,k'}u(\z)-\epsilon \overline \L_{kk'}[u](\z)+O(\epsilon^2),
\end{eqnarray}
where
\begin{eqnarray}
\fl  \L_{kk'}[u](\z)&=-D\n(\z)\cdot {\bm \nabla }_{\z}  \int_{\partial \calM}  \bigg ( {\bm \nabla }_{\y} \bigg [\sigma_{k}  \G_+(\y,s|\z) \nonumber \\
  \fl &\hspace{3cm}+kk'\sigma_{-k'} \G_+(\y,s+\alpha+\beta|\z)\bigg ]\cdot \n(\y)\bigg )u(\y) d\y  
 \end{eqnarray}
 and
 \begin{eqnarray}
\fl  \overline \L_{kk'}[u](\z)&=D\n(\z)\cdot {\bm \nabla }_{\z}  \int_{\partial \calM}  \bigg ( {\bm \nabla }_{\y} \bigg [\sigma_{k}  \G_-(\y,s|\z) \nonumber \\
  \fl &\hspace{3cm}+kk'\sigma_{-k'} \G_-(\y,s+\alpha+\beta|\z)\bigg ]\cdot \n(\y)\bigg )u(\y) d\y  
 \end{eqnarray}
 are examples of so-called Dirichlet-to-Neumann (D-to-N) operators acting on $L^2(\partial \calM)$. (D-to-N operators also play an important role in the spectral decomposition of solutions to classical Robin boundary value problems for diffusion processes \cite{Grebenkov19}.) 
 We can now substitute equation (\ref{Jeps}) into (\ref{comQa}) and take the limit $\epsilon \rightarrow 0$ to find that
   \begin{eqnarray}
\label{comQa2}
 \L_{-1,-1}[Q^{(0)}](\z) &= \lim_{\epsilon \rightarrow 0} \frac{1}{\epsilon} H_Q^{(\epsilon)}(\z) +   \L_{1,-1}[\widetilde{P}^{(0)}](\z) .
\end{eqnarray}
The right-hand side exists as $H_Q^{(\epsilon)}(\z)=O(\epsilon)$ due to the Dirichlet boundary condition imposed on $\partial \calM$. Similarly, from equations (\ref{comRa}) and (\ref{Keps}), we have
  \begin{eqnarray}
\label{comQa2}
 \overline \L_{-1,-1}[R^{(0)}](\z) &= \lim_{\epsilon \rightarrow 0} \frac{1}{\epsilon}  H_R^{(\epsilon)}(\z) + \overline\L_{1,-1}[\widetilde{P}^{(0)}](\z) .
\end{eqnarray}
Finally,
\begin{eqnarray}
\fl & \L_{1,1}[\PP^{(0)}](\z) +\overline\L_{1,1}[\widetilde{P}^{(0)}](\z) = \lim_{\epsilon \rightarrow 0} \frac{1}{\epsilon}  H^{(\epsilon)}_P(\z)+ \frac{1}{2}\bigg (\L_{-1,1}[\Q^{(0)}](\z) + \overline\L_{-1,1}[\widetilde{R}^{(0)}](\z) \bigg ) .\nonumber \\
\fl
  \end{eqnarray}
  This establishes the existence of the solution to the renewal equations (\ref{mrenewala}) and (\ref{mrenewalb}) in the limit $\epsilon \rightarrow 0$. However, obtaining an explicit solution using the spectral properties of the D-to-N operators is nontrivial. First, the exact spectrum of a D-to-N operator is only known for special geometries such as the interior of exterior of a sphere \cite{Grebenkov19}. Second, the Fredholm integral equations involve several linearly independent operators so that any spectral decomposition results in an infinite-dimensional matrix equation.
 
\section{Discussion}

 In this paper we developed a renewal theory for single-particle diffusion across a stochastically-gated interface. We assumed that the gate randomly switches between a closed (impermeable) and an open (permeable) state. We constructed a renewal equation that relates the joint probability density for particle position and the state of the gate to the marginal probability density and FPT density for BM restricted to one or other side of the interface, each of which is treated as a totally absorbing boundary. Following each round of absorption, BM is immediately restarted at a perpendicular distance $\epsilon$ from the interface. In the 1D case we explicitly solved the Laplace transformed renewal equation for $\epsilon >0$, which reduced to the solution of a corresponding forward Kolmogorov equation in the limit $\epsilon \rightarrow 0$. We then showed how extending the analysis to a higher-dimensional interface $\partial \calM$ leads to a system of Fredholm integral equations for functions on $L^2(\partial \calM)$. We suggested one method for solving these equations, namely, using a spectral decomposition based on eigenfunctions of linear operators on $L^2(\partial \calM)$ that reduce to D-to-N operators in the limit $\epsilon \rightarrow 0$.
 
In order to develop the basic mathematical framework, we restricted our analysis to (i) a single interface and (ii) a two-state Markov chain model of a switching gate. One natural extension of the former would be to consider 3D diffusion in a multilayered medium consisting of an array of parallel planar interfaces. Assuming that diffusion rapidly equilibriates in the lateral directions, we could then reduce the system to an effective 1D model with multiple stochastically gated interfaces, and then develop a renewal formulation along analogous lines to previous work on snapping out BM \cite{Bressloff23b}. However, one additional level of complexity is that for $N$ independently switching gates there would be $2^N$ different gate configurations. Another non-trivial extension would be to consider more general non-Markovian models of a switching gate. However, a major advantage of Markovian switching is that the transition matrix $\Pi_{kk_0}(t)$ consists of a linear combination of exponential functions of time, see equation (\ref{Pi}). This allowed us to take Laplace transforms of products such as $p_{k|k_0}(x,t|x_0)=\Pi_{kk_0}(t)p(x,t|x_0)$, see equation (\ref{pk}), which would no longer be possible for non-Markovian switching.

Finally, combined with our previous work on snapping out BM \cite{Bressloff22,Bressloff23a,Bressloff23b}, renewal theory provides a general mathematical framework for modelling diffusion through semipermeable and stochastically gated interfaces. The core advantage compared to boundary value problems based on Kolmogorov equations is that there is an explicit separation between the first passage time problem of detecting the gated interface (absorption) and the subsequent rule for restarting the BM. This separation also better reflects the probabilistic structure of sample paths realised by a Brownian particle in the presence of an interface. However, finding analytical solutions of the resulting renewal equations is mathematically challenging, particularly in the case of multiple and spatially extended interfaces. Clearly, in future work it will be necessary to develop efficient numerical and approximation schemes for obtaining such solutions.

\end{document}